\documentclass[a4paper,12pt,twoside,english]{article}
\usepackage[latin1]{inputenc}
\usepackage{parskip}               % suppress hanging indentation
\usepackage[dvips]{graphicx,epsfig,xcolor}
\usepackage{wrapfig,rotating}
\usepackage{array}
\usepackage{hyperref}
\usepackage{breakurl} % dvips suppresses url linebreaks, this has a workaround 
\usepackage{amsfonts}       % AMS fonts
\usepackage{amsmath}        % AMS maths
\usepackage{amssymb}        % AMS symbols
\usepackage{hhline}
\usepackage{feynmp}
\usepackage{multirow}
\usepackage[,hang,labelfont=bf]{caption}
\usepackage{subcaption}
\usepackage{xspace} % used by all the unit declarations

\usepackage{cite}                  % nice citations (shortens stuff)
\newlength{\dinwidth}
\newlength{\dinmargin}
\newcommand{\percent}{\,\%\xspace}

\newcommand{\NU}[2]{\ensuremath{#1\,\mathrm{#2}}\xspace}

\newcommand{\comment}[1]{}

\newcommand{\deltaP}{\ensuremath{\delta \mathcal{P} / \mathcal{P}}}
\newcommand{\Pe}{\ensuremath{\mathcal{P}_{e^-}}}
\newcommand{\Pp}{\ensuremath{\mathcal{P}_{e^+}}}
\newcommand{\Pol}{\ensuremath{\mathcal{P}}}
\newcommand{\AP}{\ensuremath{\mathcal{AP}}}
\newcommand{\A}{\ensuremath{\mathcal{A}}}
\newcommand{\paveip}{\ensuremath{{\langle\Pol\rangle_{\text{lumi}}}}}

\newcommand{\doi}[1]{\href{http://dx.doi.org/#1}{doi:#1}\xspace}

\newcommand{\mb}{\mathrm{\,mb}}

\newcommand{\MGy}{\mathrm{\,MGy}}

\newcommand{\GeV}{\mathrm{\,GeV}}

\newcommand{\m}{\mathrm{\,m}}
\newcommand{\nm}{\mathrm{\,nm}}
\newcommand{\cm}{\mathrm{\,cm}}
\newcommand{\mm}{\mathrm{\,mm}}
\newcommand{\mum}{\mathrm{\,\mu m}}

\newcommand{\kHz}{\mathrm{\,kHz}}

\newcommand{\ms}{\mathrm{\,ms}}

\newcommand{\fs}{\mathrm{\,fs}}

\newcommand{\ns}{\mathrm{\,ns}}

\newcommand{\T}{\mathrm{\,T}}
\newcommand{\D}{\mathrm{\,D}}

\newcommand{\mrad}{\mathrm{\,mrad}}

\newcommand{\fC}{\mathrm{\,fC}}

\newcommand{\V}{\mathrm{\,V}}

\DeclareMathOperator{\sgn}{sgn}

\DeclareMathOperator{\Corr}{Corr}

\newcommand{\approptoinn}[2]{\mathrel{\vcenter{
  \offinterlineskip\halign{\hfil$##$\cr
    #1\propto\cr\noalign{\kern2pt}#1\sim\cr\noalign{\kern-2pt}}}}}

\newcommand{\appropto}{\mathpalette\approptoinn\relax}
\makeatletter
\@ifundefined{subfloat}{\newcommand{\subfloat}}{}
\makeatother
\renewcommand{\subfloat}[3][.45\linewidth]{ \begin{subfigure}[b]{#1}
    \centering
    \includegraphics[width=0.95\textwidth]{#2}
    \caption{}
    \label{#3}
  \end{subfigure}}

\begin{document}
%% =================================================================
%% =====  Title Page  ==============================================
%% =================================================================
\begin{titlepage}
  \begin{flushleft}
    {\tt DESY 15-170    \hfill    ISSN 0418-9833} \\
    %{\tt LC-DET-2015-xxx}                \\
    {\tt September 2015}                  \\
  \end{flushleft}

  \vspace{1.0cm}
  \begin{center}
    \begin{Large}
      {\bfseries \boldmath A Calibration System for Compton Polarimetry at $e^+e^-$ Linear Colliders}

      \vspace{1.5cm}
      Benedikt Vormwald$^{1,2}$ , Jenny List$^1$, and Annika Vauth$^{1,2}$
    \end{Large}

    \vspace{.3cm}
    % Addresses and Institutions  (remove "1- " in case of a single institution)
    1- Deutsches Elektronen-Synchrotron DESY \\ 
    Notkestr. 85,  22607 Hamburg, Germany
    \vspace{.1cm} \\
    % ------------------------------------
    2- Universit\"at Hamburg, Institut f\"ur Experimentalphysik \\
    Luruper Chaussee 149,  22761 Hamburg, Germany
  \end{center}

  \vspace{1cm}

  \begin{abstract}
Polarimetry with permille-level precision is essential for future electron-positron linear colliders. Compton polarimeters can reach negligible statistical uncertainties within seconds of measurement time. The dominating systematic uncertainties originate from the response and
alignment of the detector which records the Compton scattered electrons. The robust baseline technology for the Compton polarimeters foreseen at future linear colliders is based on an array of gas Cherenkov detectors read out by photomultipliers.
In this paper, we will present a calibration method which promises to monitor
nonlinearities in the response of such a detector at the level of a few permille.
This method has been implemented in an LED-based calibration system which matches the existing prototype detector.
The performance of this calibration system is sufficient to control the corresponding contribution to the total uncertainty on the extracted polarisation to better than $0.1\%$.

  \end{abstract}

\end{titlepage}

% \begin{fmffile}{fgraphs}
% =============================================================================
\section{Introduction}        \label{sec:intro}
Since the event rates of electroweak processes depend on the chirality of the colliding particles, beam polarisation is a key ingredient of the physics programme of
future electron-positron colliders~\cite{bib:power} and
its precise knowledge is as important as the knowledge of the luminosity. 
Thereby, the luminosity-weighted average polarisations at the interaction point \paveip, which is the value relevant for the interpretation of collider data, have to be
determined by combining fast measurements of Compton polarimeters with long-term scale calibration 
obtained from reference processes in collision data. 
 
In particular for the International Linear Collider~\cite{tdr}, where both beams are foreseen to be 
longitudinally polarised, it is required to control \paveip\ with per mille-level precision in order not to limit the precision of cross-section measurements. 
Two Compton polarimeters~\cite{boogert} per beam aim to measure the longitudinal polarisation 
before and after the collision with a precision of $\deltaP \leq 0.25\%$\footnote{for typical 
ILC beam polarisation values of $\Pe \geq 80\%$ and $\Pp \geq 30\%$ or even $\geq 60\%$.}. 
It should be noted, though, that this goal is not driven by physics requirements, but by what 
used to be considered feasible experimentally. Thus, further improvements in polarimetry would 
still have direct benefits for the physics potential of the machine.

In order to evaluate the effects on the mean polarisation vector 
caused by the beamline magnets between the polarimeters and the interaction point, 
by the detector magnets and by the beam-beam interaction, spin tracking simulations are required. These effects have recently been studied in~\cite{spintracking}, concluding that not only a cross-calibration 
of the polarimeters to 0.1\percent is feasible, but also individual extrapolations of upstream and downstream measurements to the $e^{+}e^{-}$ interaction point.

From reference $e^+e^-$ reactions, the long-term average of the polarisation at the interaction point can be determined. In particular $W$ pair production~\cite{List:EPSproceedings, diss:marchesini, Rosca:2013lcnote} and single $W$ production~\cite{Graham} have been studied in this respect, showing that
precisions of about $0.15\%$ can be achieved after several years of data taking. These results are quite robust with regard to systematic uncertainties: e.g.\ 
in the measurement using $W$ pair production, even for a conservative assumption of 0.5\percent uncertainty on the selection efficiency for the signal and 5\percent for the background, the impact on the uncertainty of the polarisation measurement was found to be below the statistical uncertainty for an integrated luminosity  of $\NU{500}{fb^{-1}}$~\cite{diss:marchesini}.
Any imperfection in the beam helicity reversal, i.e.\ differences in the magnitude of the polarisation 
between measurements with left- and right-handed polarised beams, has to be corrected for based on the 
polarimeter measurements. 

The two polarimeters per beam are located about $1.8$\,km upstream and $160$\,m downstream of the 
interaction point, providing non-destructive measurements of the longitudinal beam polarisation based on the polarisation dependence of Compton scattering. They have been designed for operation at beam 
energies between $45$\,GeV and $500$\,GeV. 
Several options for detecting the Compton-scattered electrons\footnote{or positrons in case of the 
positron beam of the ILC which is equipped analogously. Throughout this paper, we'll use ``electron'' to refer to both cases, unless explicitly stated otherwise.} are being considered. 
The baseline solution is a gas Cherenkov detector consisting 
of $20$ channels, each $1$\,cm wide. A two-channel prototype of such a detector has been 
successfully operated in testbeam, where the alignment requirement for the ILC 
has nearly been reached and electronic noise was found to be at a negligible level~\cite{testbox_paper}. 
The remaining challenge for per mille-level polarimetry with a gas Cherenkov detector is the
linearity of the device.

In this paper, we present a calibration system which is able to monitor nonlinearities with sufficient
precision and provides the abilities to correct for them. The paper is organised in three main parts: In section~\ref{sec:poldetreq}, we summarise the most important facts about the ILC Compton polarimeters
before introducing the baseline detector for the Compton scattered electrons and discussing the systematic uncertainties of the measurement,
which define the target precision of the calibration system.
In section~\ref{sec:led}, we introduce a suitable linearisation method, discuss the resulting requirements,
and present a realisation of a suitable light source.  The 
performance of the calibration system and its application to the ILC Compton polarimeters will be discussed in section~\ref{sec:meas_and_ILC}
before concluding in section~\ref{sec:conclusion}.

% % =============================================================================
%###############################################################################
\section{Compton Polarimetry at the ILC} 
\label{sec:poldetreq}
%###############################################################################
Compton scattering of circularly polarised laser light off the lepton beam is the 
method of choice for polarimetry at high-energy colliders for a number of 
reasons: As a pure QED process, the scattering cross-section can be calculated 
precisely, with radiative corrections less than $0.1\%$~\cite{Swartz:1997im}, 
and offers large polarisation asymmetries. The interaction with the laser 
conserves the beam quality so that polarimetry does not interfere with collision 
data taking. The laser polarisation can be flipped very quickly using e.g.\ a 
Pockel's cell and its absolute value can be controlled at the level of 
$0.1\%$~\cite{Abe:2000dq}. Backgrounds will be
determined in-situ by shutting the laser off and thus can easily be corrected 
for.

Compton polarimetry was successfully employed at many previous accelerators, for 
instance at the SLC~\cite{Abe:2000dq} or at HERA~\cite{Barber:1992fc, Barber:1994ew}.
In this section, we will summarise the most important generic aspects of Compton 
polarimetry following the notation from~\cite{teslanote}, before summarising the 
current design of the polarimeters foreseen at the ILC. Additional details on 
the ILC polarimeters can be found in~\cite{boogert}.

%*******************************************************************************
\subsection{Compton Polarimetry Basics}
\label{subsec:polarimetry}
%*******************************************************************************

%-------------------------------------------------------------------------------
%\subsubsection{Compton Polarimetry Basics}
%\label{subsubsec:polBasics}
%-------------------------------------------------------------------------------
The kinematics of Compton scattering are fully determined by two variables, e.g.\ the 
azimuthal and polar angle of the scattered electron. In particular, the polar 
angle is equivalent to the energy of the scattered electron. 
Besides the unpolarised part, the double-differential Compton cross section~\cite{fano} 
contains terms proportional to the linear light polarisation and to the product
of circular light polarisation and transverse or longitudinal electron 
polarisation.  When integrating the cross section over the full azimuthal angle, only the term
containing the longitudinal polarisation remains, so that any sensitivity to
 linear light polarisation and  transverse electron polarisation is removed. 
Thus transverse beam polarisation requires a double differential measurement
in energy and e.g.\ vertical position\footnote{For a recent discussion of 
transverse polarimetry at the ILC c.f.~\cite{etai}.}, while for longitudinal 
polarimetry, which is the primary concern at the ILC, it is sufficient to
measure either the energy of the scattered electron or the scattered photon.

The minimum energy of the electrons
is given by the Compton edge energy
\begin{equation}
E_{\mathrm{min}} = E_0 \frac{1}{1+x} \quad,
\end{equation} 
where $E_0$ is the beam energy, while the dimensionless variable $x$ relates 
$E_0$ to the energy of the laser photons $\omega_0$, the electron mass $m$ and 
the crossing angle $\theta_0$ between the electron beam and the laser:
\begin{equation}
x = \frac{4E_0\omega_0}{m^2} \cos^2{(\theta_0/2)} \simeq 
\frac{4E_0\omega_0}{m^2}
\end{equation} 
For e.g.\ a green laser ($\omega_0 = 2.33$\,eV) shining head-on to an electron 
beam with $E_0 = 250$\,GeV, $x\simeq 9$ and thus $E_{\mathrm{min}} \simeq 0.1 
E_0 = 25$\,GeV, whereas the remaining energy is carried away by the photon.  

In the laboratory frame, both the scattered electrons and photons are contained 
in a narrow forward cone around the electron beam direction, with typical 
opening angles of $10-20\mu$rad.

The differential Compton scattering cross section with respect to the normalised 
photon energy $y=\omega/E_0$ can be written as~\cite{fano} 
\begin{equation}
 \frac{\mathrm{d}\sigma(\lambda,\mathcal{P}_z)}{\mathrm{d}y} = 
\frac{2\sigma_0}{x}\left[\frac{1}{1-y}+1-y-4r(1-r)+\mathcal{P}_z\lambda 
rx(1-2r)(2-y)\right] \label{eq:differentialcomptonxsec}
\end{equation}
or with respect to the absolute energy of the Compton scattered 
electron it becomes
\begin{equation}
 \frac{\mathrm{d}\sigma}{\mathrm{d}E} = \frac{1}{E_0} 
\frac{\mathrm{d}\sigma}{\mathrm{d}y},
\end{equation}
where 
\begin{eqnarray}
\sigma_0 = 249.5\mb \qquad \text{and} \qquad r =\frac{y}{x(1-y)}.
\end{eqnarray}
$\mathcal{P}_z$ is the longitudinal electron beam polarisation and $\lambda$ 
describes the laser polarisation.
Thus, the differential cross section is separated into a spin-independent and 
spin-dependent part. For convenience of the reader we note that $r=1$
corresponds to the Compton edge.

From equation~\ref{eq:differentialcomptonxsec} it can be seen that the asymmetry between the differential cross sections corresponding to the two laser helicity states $\sgn{\lambda}=\pm 1$ is proportional to the product of laser and beam polarisation:
\begin{eqnarray}
 \A(\Pol_z) 
 &=& \frac{d\sigma^-(\Pol_z) - d\sigma^+(\Pol_z)} 
        {d\sigma^-(\Pol_z) + d\sigma^+(\Pol_z)}
\propto   |\lambda \Pol_z|
\label{eq:asymmetry_sigma}
\end{eqnarray}
Herein, $\sigma^+$ ($\sigma^-$) denotes the case of parallel (anti-parallel) laser and electron helicities.   The proportionality factor is given by the inverse of the analysing power \AP, which is the asymmetry expected for $\lambda \Pol_z = 100\percent$.
 
\begin{figure}[htb]
\centering
 \begin{subfigure}[b]{.48\linewidth}
  \centering
  \includegraphics[width=0.97\textwidth]{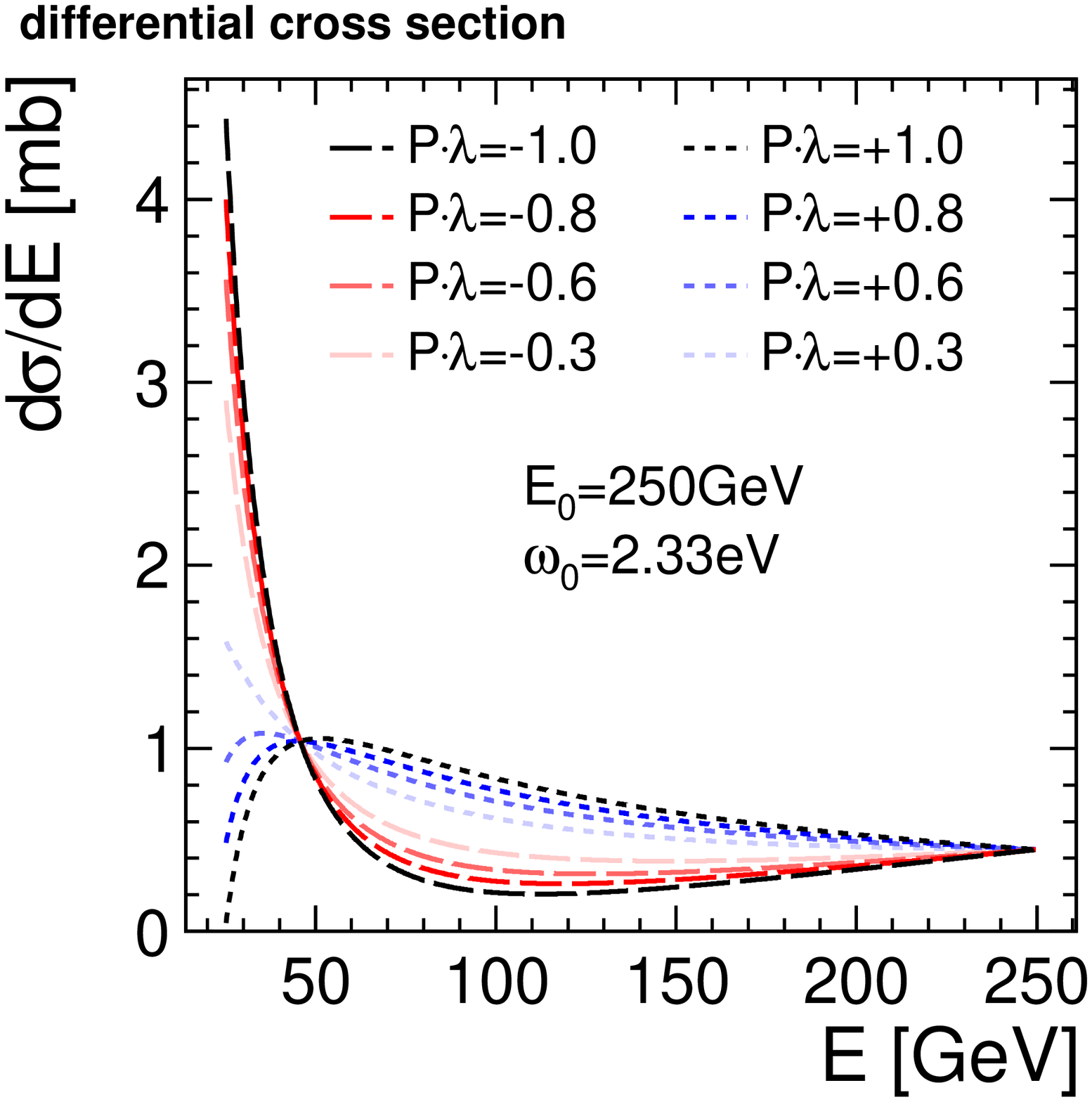}
  \caption{}
  \label{fig:req:xsec:differential}
 \end{subfigure}
 \begin{subfigure}[b]{.48\linewidth}
  \centering
  \includegraphics[width=0.97\textwidth]{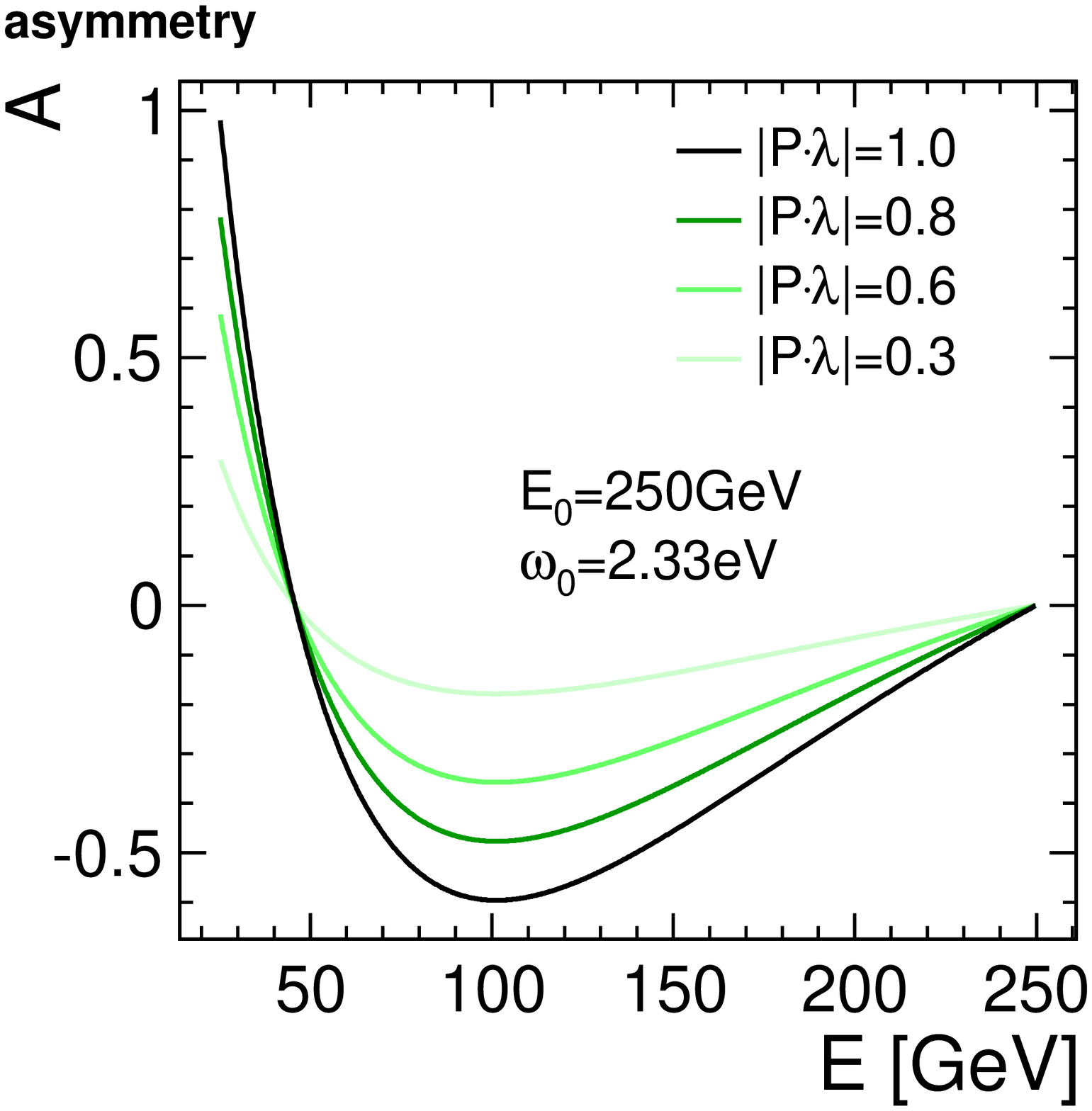}
  \caption{}
  \label{fig:req:xsec:asymmetry}
 \end{subfigure}
 \caption{(\subref{fig:req:xsec:differential}) Differential Compton scattering 
cross section for a beam energy of $E_0=250\GeV$ and green laser light. The 
different colours show the dependence of the differential cross section on the 
product of laser and beam polarisation.
          (\subref{fig:req:xsec:asymmetry}) Asymmetry for different absolute values of the beam 
polarisation. The asymmetry corresponding to $|\mathcal{P}_z\cdot\lambda|=1.0$ 
(black) is also called analysing power.}
 \label{fig:req:xsec}
\end{figure}

Figure~\ref{fig:req:xsec} shows the differential Compton scattering cross 
section as well as the corresponding asymmetry for different beam polarisations 
with respect to the energy of the Compton scattered electron.
In these figures, the beam energy is set to  $E_0=250\GeV$ and a green laser is 
chosen, as this reflects the ILC baseline design parameters.

In practice, $d \sigma$ will be determined by measuring the count rate in a detector channel
with finite size, typically after converting the energy spectrum of the Compton scattered 
electrons into a position distribution via a magnetic chicane, as will be explained in the next section.
Per channel $i$, the observed asymmetry is given by
\begin{eqnarray}
\A_i &=& \frac{N_i^-(\Pol_z) - N_i^+(\Pol_z)}{N_i^-(\Pol_z) + N_i^+(\Pol_z)}
\label{eq:asymmetry_N}   
\propto \mathcal{P}_z,\label{eq:asymmetry}
\end{eqnarray}
assuming the same luminosities for both laser helicities. 

The measured polarisation values from the individual channels can then be combined
into a weighted average, where the weights $w_i$ are chosen to minimise the total statistical uncertainty, which is achieved by giving the largest weights to the channels with largest analysing power $\AP_i$~\cite{teslanote}:
\begin{equation}
\label{eq:weights}
w_i = \frac{1}{(\Pol_z \cdot \AP_i)^{-2} - 1}
\end{equation}

%-------------------------------------------------------------------------------
\subsection{The ILC Compton Polarimeters}
\label{subsec:polarimeters}
%-------------------------------------------------------------------------------
% In order to achieve a fast measurement with statistical uncertainties below 
% $1\%$ in a few seconds, operation in a multi-event mode is required with 
% $\mathcal{O}(10^3)$ Compton interactions per bunch crossing.
The ILC Compton polarimeters will be operated in a multi-event mode with 
$\mathcal{O}(10^3)$ Compton interactions per bunch crossing.
For the polarisation measurement, both the scattered photon and 
electron can be analysed equivalently.
At HERA, the longitudinal polarimeter operated successfully in a multi-photon 
mode by detecting the sum of the photon energies in a 
calorimeter~\cite{Beckmann:2000cg}.
At ILC energies and intensities, however, the cumulative energy of 
$\mathcal{O}(10^3)$ photons carrying on average $\mathcal{O}(10)$\,GeV each 
would amount to $\mathcal{O}(10)$\,TeV per bunch crossing, all within a narrow 
cone of $\mathcal{O}(10)\,\mu$rad.
Therefore, it seems highly unlikely that photon detection could contribute to 
precision polarimetry at the ILC.

The situation is very different for the detection of the scattered 
electrons as employed at SLD.
As charged particles, the electrons can be energy-analysed by dipole magnets.
This translates the energy measurement into a position measurement which does 
not require a total absorption of the energy.
Thus, the shape of the spectrum can be monitored under multi-event conditions.
Additionally, regions of high or low analysing power 
 can be evaluated separately and enter 
in the total measurement with an appropriate weight (c.f.~equation~\ref{eq:weights}).

\begin{figure}[htb]
\centering
  \includegraphics[width=0.95\linewidth]{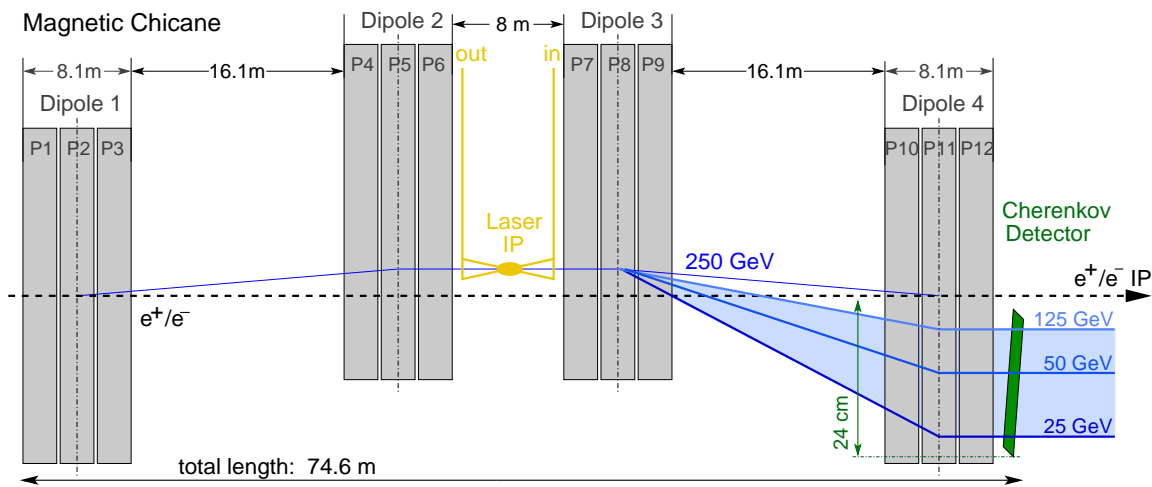}
  \caption{Sketch of the magnetic spectrometer for the ILC upstream polarimeter 
  (from~\cite{testbox_paper}). The Compton-scattered electrons with the 
  lowest energies are deflected most by the spectrometer.  }
  \label{fig:magnetic-chicane}
\end{figure}

The polarimeter chicanes consist of a sets of dipole magnets.
Figure~\ref{fig:magnetic-chicane} shows as an example a sketch of the chicane foreseen for the ILC's upstream polarimeters. The magnetic field strengths of the magnets are chosen such that the fan of 
Compton scattered electrons are deflected far enough from the main beam so that a segmented 
counting detector can be safely placed and operated next to the beam pipe. Both
the upstream and the downstream polarimeter chicanes will be operated at a constant magnetic
field for all beam energies. This implies that the Compton interaction point moves
laterally with the beam energy, but offers the advantage that the position and size of
the Compton spectrum in the detection plane stays invariant. Thus, the same measurement
quality can be maintained for all beam energies.
A detailed description of the polarimeter chicane designs for the up- and downstream 
polarimeter is discussed in reference \cite{boogert}.

%-------------------------------------------------------------------------------
\subsubsection{The Upstream Polarimeters at the ILC}
\label{subsubsec:detReqUpILC}
%-------------------------------------------------------------------------------
At the upstream polarimeter, the beam spot of the yet undisturbed bunches is small
and low background levels are expected~\cite{teslanote}.
Therefore, conventional lasers featuring moderate pulse energies can be used in order 
to probe every single bunch in an ILC bunch train. Different ILC beam parameter sets 
foresee between 1312 and 2625 bunches per bunch train,
where the trains last $1\ms$ each and occur with a repetition rate of $5$\,Hz~\cite{tdr}.
The ability to probe every bunch in a bunch train allows to accumulate high statistics
in a very short time. Furthermore, it allows to resolve possible variations in the bunch 
polarisation within a bunch train which could be present due to beam dynamics. 

On the downside, the expected radiation exposure of the detector from the Compton electrons
alone amounts to up to about $1\MGy$ per year, which requires to employ sufficiently 
radiation hard detector technologies. The resulting readout rate in the order of $10$ MHz
requires a fast responding detector as well as fast readout electronics.

The upstream polarimeter chicane as sketched in figure~\ref{fig:magnetic-chicane} spreads
the Compton spectrum over approx. $20$\,cm in the horizontal plane. Due to the benign
conditions, the instrumentation can cover nearly the whole spectrum down to about $2$\,cm 
from the undisturbed main beam, corresponding to Compton electrons up to $E=125\GeV$.
This clearly includes the zero crossing point of the asymmetry (c.f.\
figure~\ref{fig:req:xsec:asymmetry}), which together with the Compton edge position provides
important calibration reference.

%%%%%%%%%%%%%%%%%%%%%%%%%%%%%%%%%%%%%%%%%%%%%%%%%%%%%%%%%%%%%%%%%%%%%%%%%%%%%%%%%%%%%%%%%%%%

%-------------------------------------------------------------------------------
\subsubsection{The Downstream Polarimeters at the ILC}
\label{subsubsec:detReqDownILC}
%-------------------------------------------------------------------------------
The downstream polarimeter chicane starts $120\m$ behind the $e^+e^-$ interaction point, where
the Compton interaction point is located at a secondary focus point of the
machine optics. However, due to the energy loss during collisions, the beam cross section 
is four orders of magnitudes larger than at the upstream polarimeter~\cite{spintracking}. 
In addition, significant beam background at the level of ${\cal{O}}(10^3)$ photons and ${\cal{O}}(10^2)$  charged particles per bunch crossing is expected~\cite{Ken:EPWS08proceedings}.
For this reason, the laser power needs to be increased by more than three orders 
of magnitude to still maintain a reasonable signal to background 
ratio. This, in turn, reduces the maximal laser repetition rate to ${\mathcal{O}}(1)$ 
laser pulse per ILC bunch train. Therefore, the detector readout speed is of no concern.

For the same reason, the radiation dose due to the Compton electrons themselves is three 
orders of magnitude lower than at  the upstream polarimeter. On the other hand, the substantial
backgrounds, which are present for every bunch independently of the presence of a laser pulse, 
will add to the total radiation dose. Thus it seems prudent to keep similar
radiation tolerance requirements as for the upstream polarimeter. 

Since the extraction line is designed with an aperture of $0.75\mrad$ with 
respect to the interaction point, the polarimeter detector needs to be placed in 
a distance of $15\cm$ to the main beam trajectory.
In order to still be able to probe a significant part of the fan of Compton 
scattered electrons, the polarimeter chicane is slightly modified and 
two additional magnets operating at a higher field strength are inserted in the 
spectrometer part resulting in a larger dispersion \cite{Moffeit:2007tz}.
At a beam energy of $E=250\GeV$, Compton scattered electrons
from $E=44\GeV$ down to the Compton edge at $E=25\GeV$ can be detected in this 
setup. The accessible spectrum is spread over a distance of approximately $13\cm$.
Due to the expected background, detection methods insensitive to photons and with an intrinsic energy 
threshold for charged particles are considered advantageous.

%*******************************************************************************
\subsection{Detectors for the ILC Compton Polarimeters}
\label{subsec:det}
%*******************************************************************************
A natural choice for fast counting detectors are Cherenkov detectors read-out by a
photomultiplier. For relativistic electrons, this detection mechanism is intrinsically
linear, since the amount of emitted Cherenkov light is proportional to the number of electrons passing through that channel. Depending on the choice of Cherenkov medium, the threshold
for Cherenkov radiation can be in the MeV regime, which suppresses backgrounds
from low energetic electron-positron pairs from beamstrahlung. This is in particular
true for gases with refractive indices only slightly above unity. E.g.\ for $C_{4}F_{10}$ 
with $n=1.0014$, the Cherenkov threshold is at $10$\,MeV. Due to their reliability, robustness
and proven performance in the so far most precise Compton polarimeter which measured the  polarisation of the electron beam at SLC~\cite{Abe:2000dq}, gas Cherenkov detectors are
considered the default solution for the ILC polarimeters. As will be
discussed below, the detector linearity is one of the most important systematic uncertainties
of this type of detector. Therefore, we present in this paper a calibration system
which is suitable to detect and correct for nonlinearities at the sub-percent level
in-situ.

Recently, a Cherenkov detector for polarimetry based on quartz has been proposed~\cite{quartzdet}.
With a much higher light yield (due to the higher refractive index) in combination with a higher granularity, such a system would robust against nonlinearities up to the percent-level. However, this concept still
awaits a full demonstration in testbeam and a full evaluation of the robustness against backgrounds, since the Cherenkov threshold is much lower. 

Ultimate granularity would be achieved with a silicon pixel detector, eliminating the question 
of linearity completely and increasing significantly the possibilities to control the alignment.
However, in this case R\&D is still required to reach a proof-of-concept level, where the main
concern is the high local data rate in the most intense region of the Compton spectrum, followed
by radiation issues.

%*******************************************************************************
\subsubsection{A Gas Cherenkov Detector for Compton Polarimetry}
\label{subsec:gas}
\begin{figure}[htb]
 \begin{subfigure}[b]{.6\linewidth}
  \centering
  \includegraphics[width=0.97\textwidth]{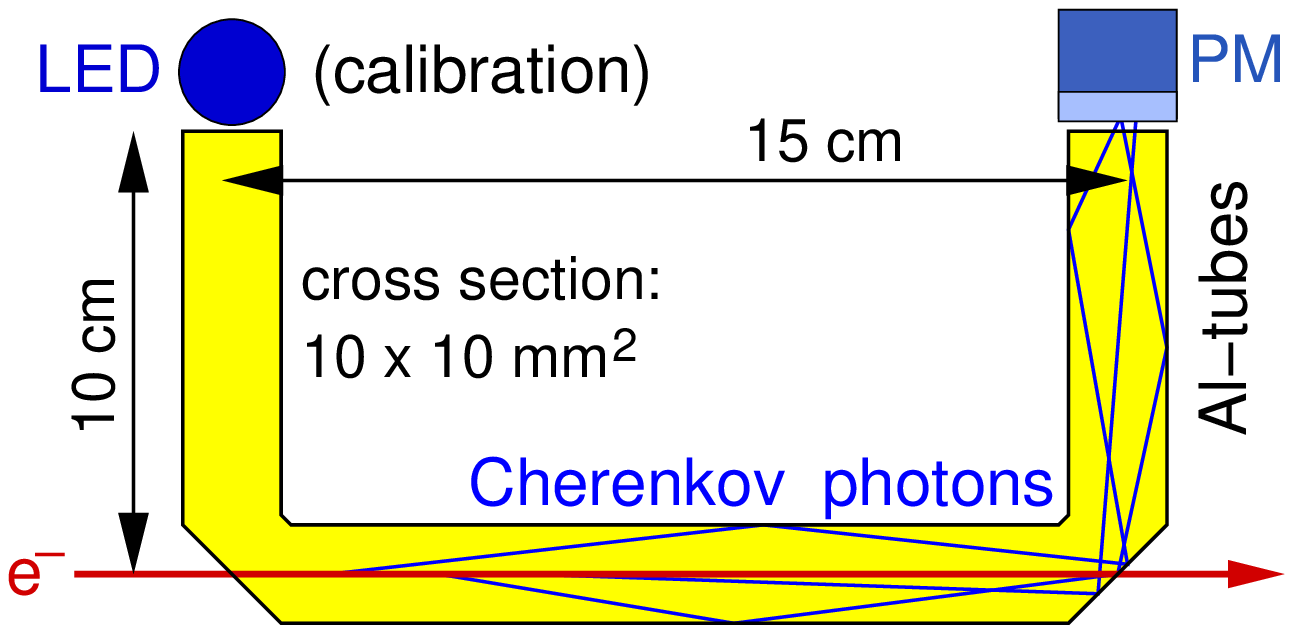}
  \caption{}
  \label{fig:gas:status:channel}
 \end{subfigure}
 \begin{subfigure}[b]{.4\linewidth}
  \centering
  \includegraphics[width=0.97\textwidth]{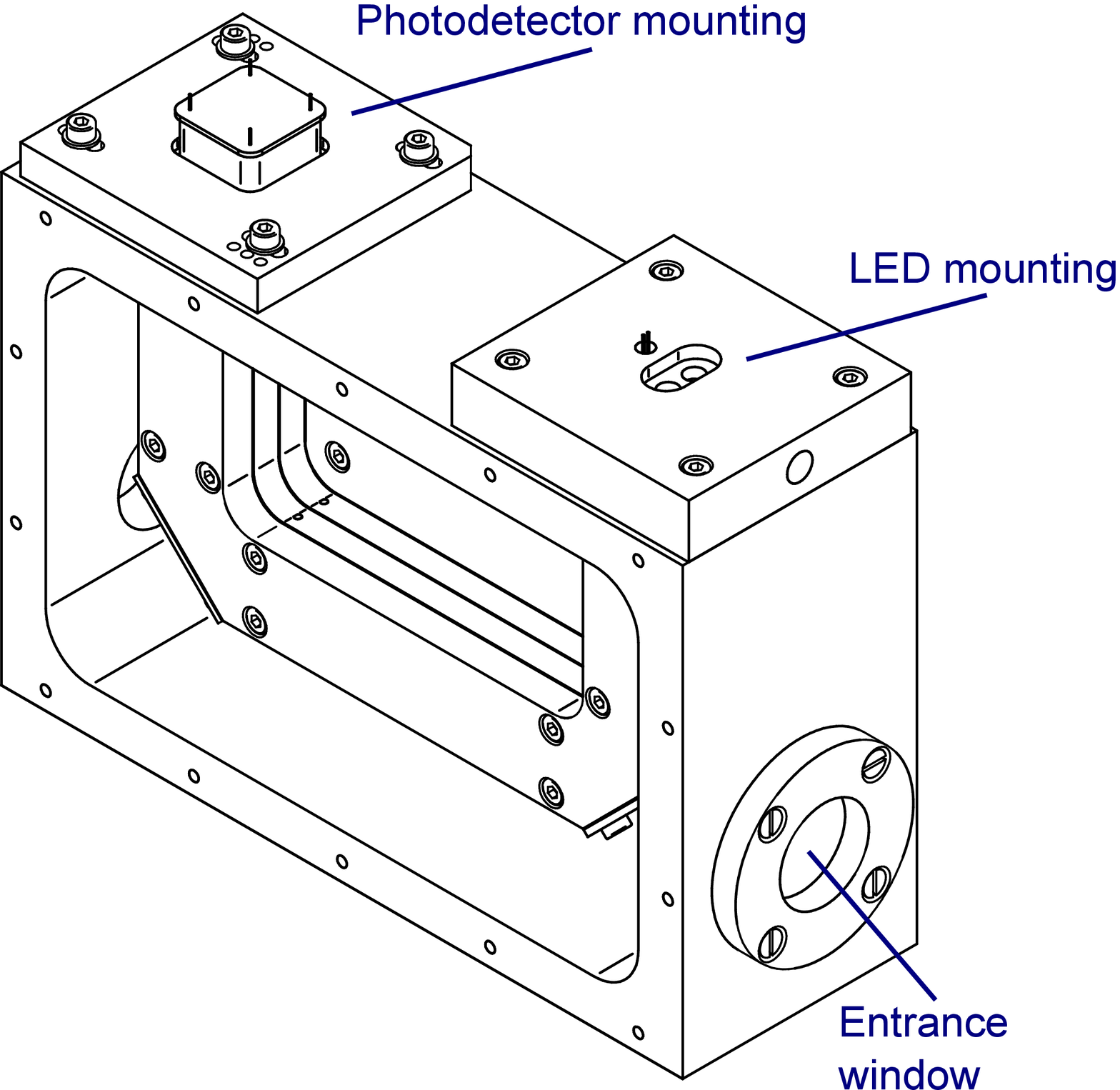}
  \caption{}
  \label{fig:gas:status:prototype}
 \end{subfigure}
 \caption{(\subref{fig:gas:status:channel}) Schematic drawing of one gas Cherenkov detector channel.
          (\subref{fig:gas:status:prototype}) Picture of the two-channel Gas Cherenkov detector prototype. Both from~\cite{testbox_paper}.}
 \label{fig:gas:status:detector}
\end{figure}

%*******************************************************************************
A prototype of a gas Cherenkov detector for ILC polarimetry has been designed, built and
operated successfully in testbeam~\cite{testbox_paper}. Based on the experience gained at SLC, the design of the channels has been improved to U-shaped aluminium tubes with a square cross section of $1\times1\cm$.
A schematic side view of one channel is depicted in Figure~\ref{fig:gas:status:channel}.
When the electrons traverse the horizontal tube of the Cherenkov counter, they induce Cherenkov light.
Mirrors at the end of the tube reflect the Cherenkov light upwards such that the actual photo detector can be placed outside the Compton scattering plane. The other upward pointing ``leg''
of the U-shaped channels is foreseen to host an LED-based calibration system. 
For the upstream polarimeter, $20$ such channels would be staggered along the shallow angle exit window.

Figure~\ref{fig:gas:status:prototype} shows a drawing of the prototype which consists out of two
non-staggered channels. In testbeam, it was demonstrated that the aimed detector alignment goals can be achieved, in particular when multi-anode photomultipliers (MAPMs) are used which allow to measure asymmetries in light distribution in each channel.
Since the light pattern on the detector plane is related to the detector--beam alignment, the alignment can also be measured and monitored during a physics run without spending dedicated beam time on alignment runs~\cite{testbox_paper}.

\begin{figure}[htb]
  \centering
  \includegraphics[width=0.6\textwidth]{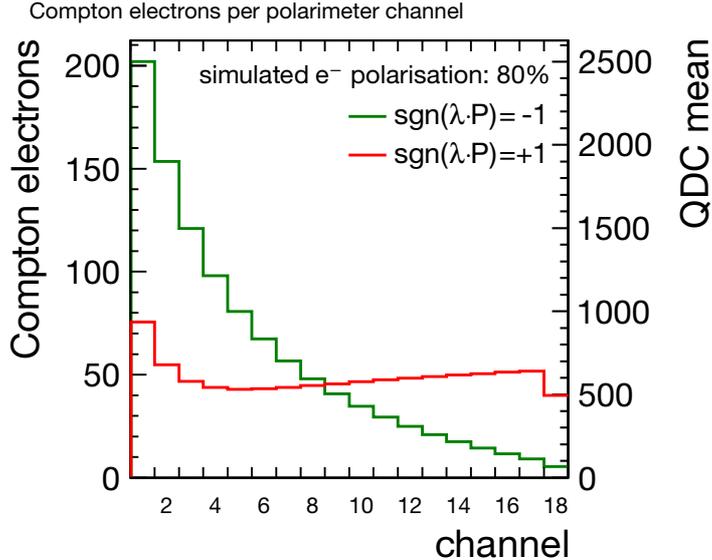}
  \caption{Simulated number of Compton scattered electrons per polarimeter 
  channel for the different laser helicities. The right axis shows the expected 
  QDC signal assuming the parameters of the prototype detector, see text.
  %a light yield of $\eta=6.5$ photoelectrons/ Compton 
  %electron \cite{testbox_paper}, a photomultiplier gain of $g=3.0\times10^5$ 
  %and a QDC resolution of $25\fC/\mathrm{QDC\,count}$.
  }
 \label{fig:gas:expectedcomptone}
\end{figure}

Figure~\ref{fig:gas:expectedcomptone} shows the number of expected Compton 
scattered electrons per polarimeter channel for a simulated polarisation of 
$\mathcal{P}=80\%$, as obtained from the fast Linear Collider Polarimeter Simulation \texttt{LCPolMC}~\cite{spintracking, Eyser2007}. The axis on the right-hand side 
illustrates the expected 
QDC signal assuming the parameters of the prototype detector~\cite{testbox_paper}, thus a 
light yield of $\eta=6.5$ photoelectrons/ Compton 
electron, a photomultiplier gain of $g=3.0\times10^5$ 
and a QDC resolution of $25\fC/\mathrm{QDC\,count}$.

%-------------------------------------------------------------------------------
\subsubsection{Sources of Systematic Uncertainties}
\label{subsubsec:systematics}
%-------------------------------------------------------------------------------
In the multi-event mode, the statistical uncertainty of the 
polarisation  measurement drops rapidly such that already after a few seconds of data taking, 
the polarisation measurement uncertainty is  dominated by systematic uncertainties.
Table~\ref{tab:polsys} summarises the envisaged error budget at the ILC in comparison to the 
obtained systematic uncertainties at the SLC polarimeter~\cite{Abe:2000dq}.

\begin{table}[hbt]
   \centering
   \begin{tabular}{lcc}
                                       & \multicolumn{2}{c}{$\deltaP $}  \\
       source of uncertainty           & SLC      & ILC goals \\
      \hline
      laser polarisation               & $0.1\%$  & $0.1\%$               \\
      detector alignment               & $0.4\%$  & $0.15 - 0.2\%$        \\
      detector linearity               & $0.2\%$  & $0.1\%$               \\
      electronic noise and beam jitter & $0.2\%$  & $0.05\%$              \\
      \hline
      Total                            & $0.5\%$  & $0.25\%$              \\
   \end{tabular}
   \caption[]{Error budget of the polarisation measurement at the ILC 
\cite{spintracking} in comparison to the determined systematic uncertainties at 
the SLC polarimeter \cite{Abe:2000dq}. }
   \label{tab:polsys}
\end{table}

\begin{description}
 \item[Laser polarisation.]
The laser polarisation and its flip accuracy have been controlled to
the level of $0.1\%$ already at the SLC polarimeter. Currently, no improvement
is envisaged for the ILC case.
\item[Noise \& jitter.]
Contributions originating from electronic noise, which were sizeable at the 
time of the SLC polarimeter, are expected to be negligible for modern DAQ 
hardware as applied in the ILC polarimeters, which is confirmed by the testbeam experience
with the prototype~\cite{testbox_paper}. Also the beam jitter is expected to be significantly 
reduced in the ILC case, since the luminosity goals require to limit the jitter to at most $10\%$ 
of the beam sizes~\cite{tdr}.
\item[Analysing power.] 
At cross-section level, the analysing power can be calculated with high precision from the Compton cross section known currently to the order $\alpha^3$~\cite{Swartz:1997im}.
Therefore, the knowledge of the analysing power is dominated by the alignment of the detector with respect
to the Compton spectrum, which includes lateral shifts of the Compton IP as well as the finite knowledge of 
the magnetic fields of the chicane. Both effects are included when the alignment is derived from
measuring the unpolarised Compton spectrum, e.g.\ from the sum of data taken with the two laser helicities.
The foreseen error budget translates into a vertical alignment precision of at 
least $100\mum$ and a precision of the tilt angle of the detector of $1\mrad$ 
with respect to the design beam orbit~\cite{quartzdet}. A granularity of about $1$\,cm
provides sufficient detail to achieve this level of control by monitoring the shape of 
Compton spectrum~\cite{testbox_paper}. It has been shown that the requirements for shifts as well as tilts 
of the detector could be met with the prototype, in particular when assisted by information of the intra-channel light pattern read-out by multi-anode PMTs~\cite{testbox_paper}.
\item[Detector nonlinearities.]
Apart from the alignment, the detector nonlinearities
are the other source of uncertainty which needs to be improved significantly in 
order to meet the precision goal for the ILC. 
From equation~\eqref{eq:asymmetry}, it becomes clear that a polarimeter 
detector does not need to be calibrated on an absolute scale, since any constant
calibration factor cancels.
However, the linearity of the detector response is essential for per mille-level 
polarimetry.

\begin{figure}[htb]
\centering
\includegraphics[width=0.53\textwidth]{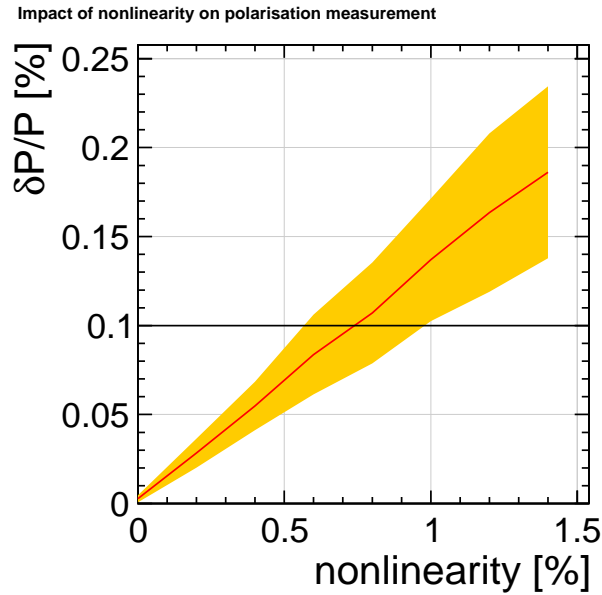}
\caption{Systematic uncertainty on polarisation measurement due to different 
detector nonlinearities. The error band shows the RMS interval obtained from 
200 randomly chosen detector response functions. The exact definition of 
nonlinearity employed here is described in section~\ref{subsec:nldefinitions}. }
%(third order polynomials) all featuring 
% the same nonlinearity. In order to keep the systematic uncertainty of the 
%polarisation measurement \deltaP\ below $<0.1\%$, 
%the detector nonlinearity must be controlled on the level of $0.5\%$.

\label{fig:gas:status:nlrequirement}
\end{figure}

In order to investigate the maximal tolerable detector nonlinearity which 
still keeps the nonlinearity contribution to the overall systematic uncertainty 
below $0.1\%$, the impact of different nonlinear detector response (or transfer) 
functions on the measurable polarisation has been 
studied in Monte Carlo simulations. For each value of the relative nonlinearity, 
200 third order polynomials with random parameters have been generated and applied 
as transfer functions in the simulated polarisation measurement.
The results are depicted in Figure~\ref{fig:gas:status:nlrequirement}.
The red line indicates the absolute value of the mean deviation between the reconstructed and 
true polarisation, while the yellow band shows the RMS of those samples.
It becomes obvious that the detector nonlinearity must be controlled at 
the level of $0.5\%$ in order to keep the resulting contribution to
\deltaP\ below $0.1\percent$. A method to achieve this is discussed in the next chapter.

\end{description}

% In order to fully exploit the advantages of beam polarisation at a future 
% linear collider, the polarisation measurement uncertainty should not 
% exceed $\delta \mathcal{P}_z / \mathcal{P}_z=0.25\%$, which is by a of factor 2 
% better than what has been achieved at the SLC polarimeter \cite{Abe:2000dq}.

% 
% % =============================================================================

%###############################################################################
\section{A Calibration System for the ILC Polarimeters} 
\label{sec:led}
%###############################################################################
In order to control the detector nonlinearity on a level of below $0.5\%$, novel
calibration techniques are needed. In this section, we present a calibration 
system, which is capable to meet these requirements.

%###############################################################################
\subsection{Definitions of Nonlinearity} 
\label{subsec:nldefinitions}
\begin{figure}[htb]
 \begin{subfigure}[b]{.48\linewidth}
  \centering
  \includegraphics[width=0.97\textwidth]{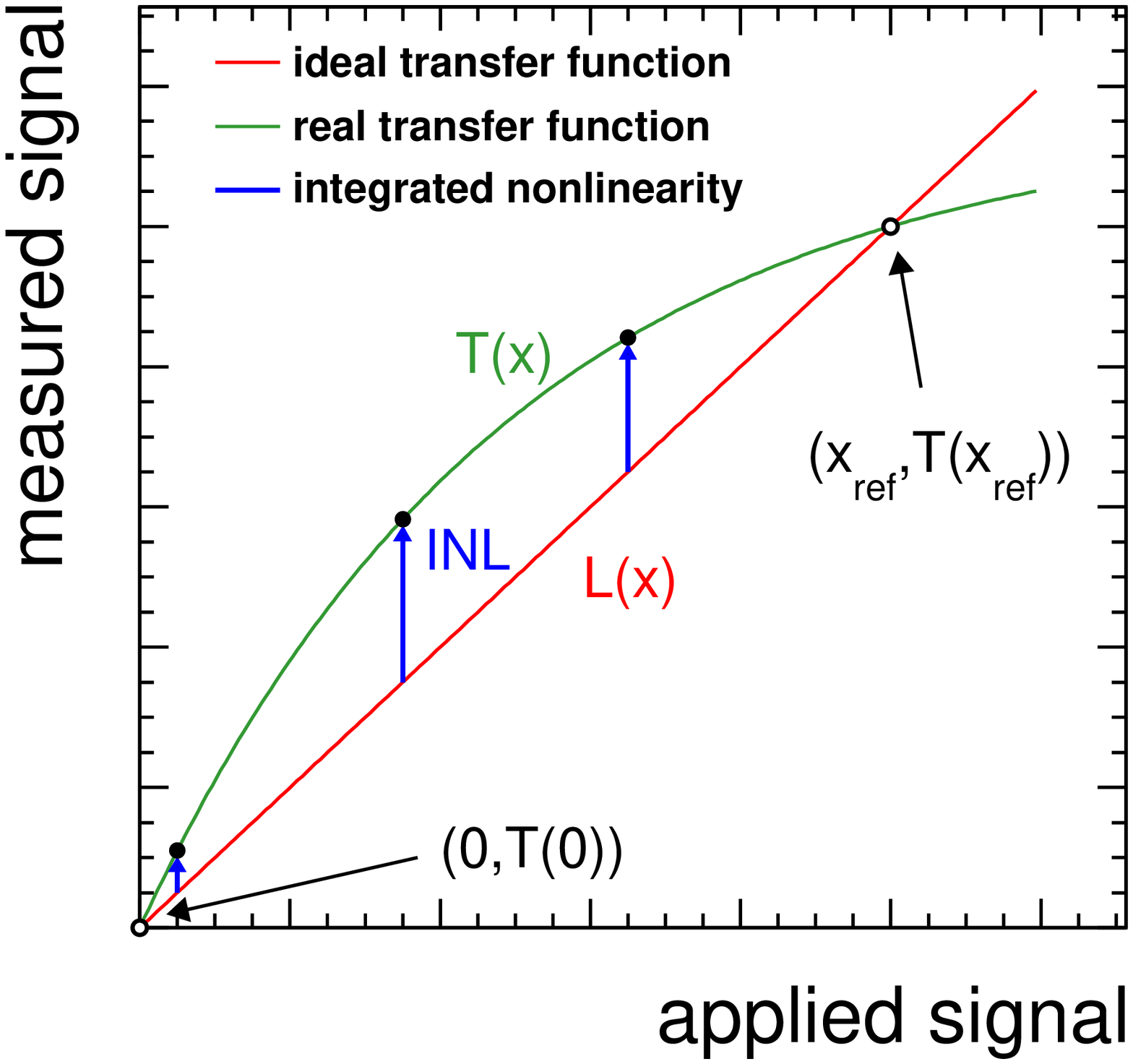}
  \caption{Integrated nonlinearity}
  \label{fig:calib:inl}
 \end{subfigure}
 \begin{subfigure}[b]{.48\linewidth}
  \centering
  \includegraphics[width=0.97\textwidth]{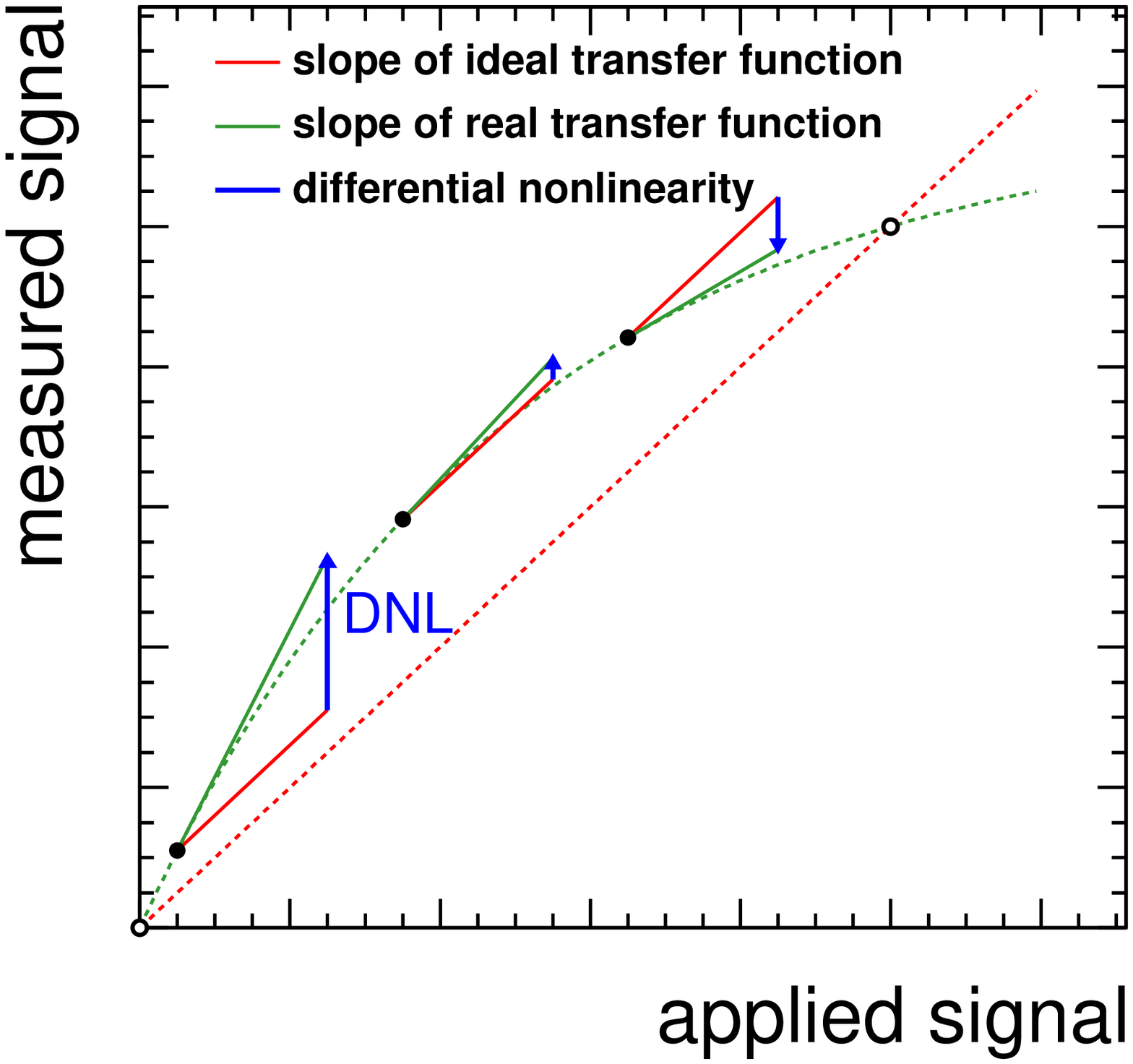}
  \caption{Differential nonlinearity}
  \label{fig:calib:dnl}
 \end{subfigure}
 \caption{Different ways of specifying the nonlinearity of a detector response.}
 \label{fig:calib:nldefinitions}
\end{figure}

%###############################################################################
%here goes INL and DNL: there are different ways of specifying a nonlinearity..."
Basically, there are two ways of describing the nonlinearity of a detector, 
which are illustrated in Figure \ref{fig:calib:nldefinitions}: The integrated 
nonlinearity (INL) specifies the absolute deviation of the detector response 
$\T(x)$ from a linear response $\mathrm{L}(x)$, where $x$ is an applied signal (c.f. 
Figure \ref{fig:calib:inl}). Thereby, the linear response is defined by the 
response to a signal $x_\mathrm{ref}$ which marks the end of the range to be 
calibrated. 
%Therefore it holds that $\T(x_\mathrm{ref})=\L(x_\mathrm{ref})$.

Another way of describing a nonlinearity is the differential nonlinearity 
(DNL). Here, the deviation between the slope of the detector transfer function 
and the linear response is used (c.f. Figure \ref{fig:calib:dnl}).
Although the DNL does not preserve any information about the absolute scale of the 
detector response, all the information regarding the linearity of the response 
is still present. Thus, also a DNL measurement can be used to linearise the 
detector response.

%It is important to note that no absolute calibration is necessary, since any 
%absolute calibration factor cancels out in equation~\eqref{eq:asymmetry}.

%It should be pointed out that the correction algorithm fails for too large 
%detector nonlinearities, since it relies on a small-nonlinearity approximation.
%In order to be able to correct nonlinearities to the anticipated level of 
%$0.5\%$, Monte-Carlo studies have shown that the relative nonlinearity should 
%not exceed $3\%$. This assumption, however, is very well fulfilled for 
%photomultipliers, which are intrinsically very linear devices.

%###############################################################################
\subsection{Permille-Level Linearisation} 
\label{subsec:dnl}

Measuring the nonlinearity of a light detector to permille-level is very 
challenging as this usually requires an intensity calibrated light source of the
same precision.
Different calibration methods based on LED light pulses were studied in 
Reference \cite{diss:helebrant}. In the following, we will 
focus on a double pulse method, which pursues a differential calibration 
approach:
One light pulse (base pulse $x$) is tunable in its light intensity over the expected 
dynamic range of the detector. A second light pulse (differential pulse 
${\Delta}x$) features a constant intensity, which is small compared to the dynamic range. 
Measuring the detector response to $x$ and $x+{\Delta}x$, allows to 
estimate the DNL. This approach is advantages as it does not require a 
calibrated light source at all as it will be demonstrated in the next section. Details on 
this method are also discussed in Reference \cite{diss:vormwald}.

%###############################################################################
\subsubsection{Mathematical Foundations and Linearisation Method}
\label{subsec:dnl:math}
In the following, we outline the mathematical foundations and the procedure of 
permille-level linearisation using the double pulse method.

% Measuring the nonlinearity of a light detector to permille-level is very 
% challenging as this usually requires an intensity calibrated light source of the
% same precision. In order to linearise a photomultiplier response, as it is needed 
% for the precise polarisation measurement, it is sufficient to only measure the 
% derivative of the detector transfer function.
% % This approach makes use of two light pulses in order to measure the derivative of the detector transfer function.
% % The first LED light pulse is tunable in its intensity whereas the light intensity of the second LED light pulse is constant and small compared to the base pulse.
% % For a perfect linear detector, the measureable difference between the detector response between the base+differential pulse and the pure base pulse should be constant.
% % Any deviation from this constant with respect to the chosen base pulse can be attributed to the detector nonlinearity and can be used for obtaining a detector specific correction function.
% This approach is advantages as it does not require a calibrated light source at all.

The most general transfer function of a detector can be parametrised as
\begin{equation}
\label{eq:transfer}
 \T(x) = A + \left( B + nl(x)\right)x.
\end{equation}
% where $\T(x)$ is the detector response to an input light signal $x$.
Herein, $A$ denotes an offset (e.g. due to dark current), which can be easily 
measured and subtracted. Thus, we set $A=0$ for simplicity. $nl(x)$ is a 
arbitrary function which describes the nonlinear part of the detector response. 
In order to linearise the detector response, the correction function
\begin{equation}
\label{eq:corr}
 \Corr(x) = \frac{B}{B + nl(x)}
\end{equation}
needs to be determined.

The derivative of $\T(x)$ is proportional to the difference between the values 
of the detector response for a dedicated light signal $x$ and the signal which 
is increased by a constant and small additional quantity of light from the 
differential light pulse ${\Delta}x$:
\begin{equation}
 \label{eq:derivative}
 \D(x) := \frac{\mathrm{d}}{\mathrm{d}x} \T(x) \approx \frac{\Delta\T(x)}{{\Delta}x}  \propto \Delta\T(x) = \T(x+{\Delta}x) - \T(x)
\end{equation}

If $nl(x)$ is small, which is intrinsically the case for photomultipliers, it 
holds
\begin{equation}
\label{eq:smallnlapprox}
 \D(x) \appropto \D(\T(x))  \appropto  \Delta\T(\T(x)).
\end{equation}
Thus, measuring $\Delta\T$ in dependence of $\T(x)$ contains (up to an 
proportionality factor) approximately the same information as the direct 
derivative of the transfer function in dependence of the absolute amount of 
injected light $x$.
% \begin{equation}
%  \left(\T(x), \Delta\T(x) \right) \appropto  \left(x, \D(x) \right) 
% \end{equation}
Thus, we achieve an approximate determination of the derivative of the transfer 
function which does not rely on the knowledge of the injected quantity of light 
$x$, but only on the corresponding detector responses.

The measurement of the data pairs $(\mathrm{T}(x),\Delta\T(\T(x)))$ can now be 
used to build the correction function $\Corr(x)$. For this purpose, the data points
are fitted with a higher order polynomial in order to obtain a continuous 
function which can then be processed further.

Combining Equations \eqref{eq:transfer} and 
\eqref{eq:derivative} gives
\begin{equation}
 \label{eq:deltat}
 \Delta\T(x) \approx \Delta x \cdot \frac{\mathrm{d}}{\mathrm{d}x} \T(x) = \Delta x \cdot \left( B + nl(x) + nl'(x)x \right).
\end{equation}
Thus, the average of $\Delta\T$ over the whole dynamic range can be identified 
as 
\begin{equation}
 \label{eq:bpart}
 \langle \Delta\T(x) \rangle = \Delta x \cdot (B + \langle nl(x) + nl'(x)x \rangle) = \Delta x \cdot B.
\end{equation}
Note that the average over the $nl(x)$-dependent terms in 
Equation~\eqref{eq:bpart} vanishes, as by definition $nl(x_\mathrm{ref})=0$:
\begin{equation}
 \frac{1}{x_\mathrm{ref}} \int_{0}^{x_\mathrm{ref}} \left(nl(x) + nl'(x)x\right) \mathrm{d}x = \frac{1}{x_\mathrm{ref}} [nl(x) x]_{0}^{x_\mathrm{ref}} = 0 .
\end{equation}
In a next step, Equation \eqref{eq:deltat} is corrected for the constant part
\begin{equation}
 \Delta\T(x) - \Delta x \cdot B \equiv \Delta\hat{\T}(x)= \Delta x (nl(x) + nl'(x)x) 
\end{equation}
and $\Delta\hat{\T}(x)$ is integrated in order to obtain an expression for 
$nl(x)$:
\begin{equation}
 \label{eq:nlpart}
 \frac{1}{x} \int_{0}^{x} \Delta\hat{\T}(x') \mathrm{d}x' = \Delta x \cdot nl(x)
\end{equation}
% For a discrete number of measurement points, Equation \eqref{eq:nlpart} can be
% rewritten as
% \begin{equation}
%  \Delta x \cdot nl(x_j) \approx \sum_{i=0}^{j} \frac{1}{2}\left(\Delta\hat{\T}(x_i) + \Delta\hat{\T}(x_{i+1})\right) \frac{\T(x_{i+1})-\T(x_{i})}{\T(x_{j})}
% \end{equation}
Finally, Equation \eqref{eq:bpart} and \eqref{eq:nlpart} can be combined in 
order to construct the correction function of Equation \eqref{eq:corr}. The 
remaining proportionality factor $\Delta x$ in \eqref{eq:bpart} and 
\eqref{eq:nlpart} cancels. It should be stressed once more that 
$\Corr(x)\approx\Corr(\mathrm{T}(x))$ can be determined without
knowing the absolute input light signal. The only requirement is that
the additional quantity of light $\Delta x$ is constant and small compared to
the quantity of light $x$ which scans the dynamic range.

\subsubsection{Monte-Carlo Study}
\label{subsec:dnl:mc}
In order to test the presented linearisation method, a Monte-Carlo simulation 
has been developed, which takes into account the simulation of the light emission, 
attenuation, detection, and digitisation. The simulation parameters  
have been taken from datasheets of the foreseen DAQ equipment or, in case of the light
source parameters, have been adjusted in agreement with measured distributions. 
With this simulation, the performance and
limitations of the described linearisation method can systematically be studied
and parameters for a measurement setup can be optimised.

As the desired linearisation is in the sub-percent range, a sufficiently large number of 
data points $(\mathrm{T}(x),\Delta\T(\T(x)))$ with a sufficiently large 
statistic needs to be acquired. From the simulation we find that the number of scan 
steps within the dynamic range plays a small role as long as it exceeds four 
scan steps. This can be understood since nonlinear contributions of higher orders 
in the transfer function play a decreasing role and, thus, more data 
points do not strongly constrain the polynomial fit anymore. 
However, in order to ensure a good modelling of the nonlinearity by the fit
it is advantages to adjust the number of scan steps according to the time 
budget of a nonlinearity measurement.
A much stronger influence on the performance of the linearisation method has the 
number of measurements taken at one scan step since this effects the 
statistical uncertainty of the individual data points.
It has been found that in order to allow for a sub-percent linearisation at
least $3\times10^6$ measurements per scan step are desirable.

\begin{figure}[htb]
 \centering
   \begin{subfigure}[b]{.49\linewidth}
  \centering
 \includegraphics[width=0.9\textwidth]{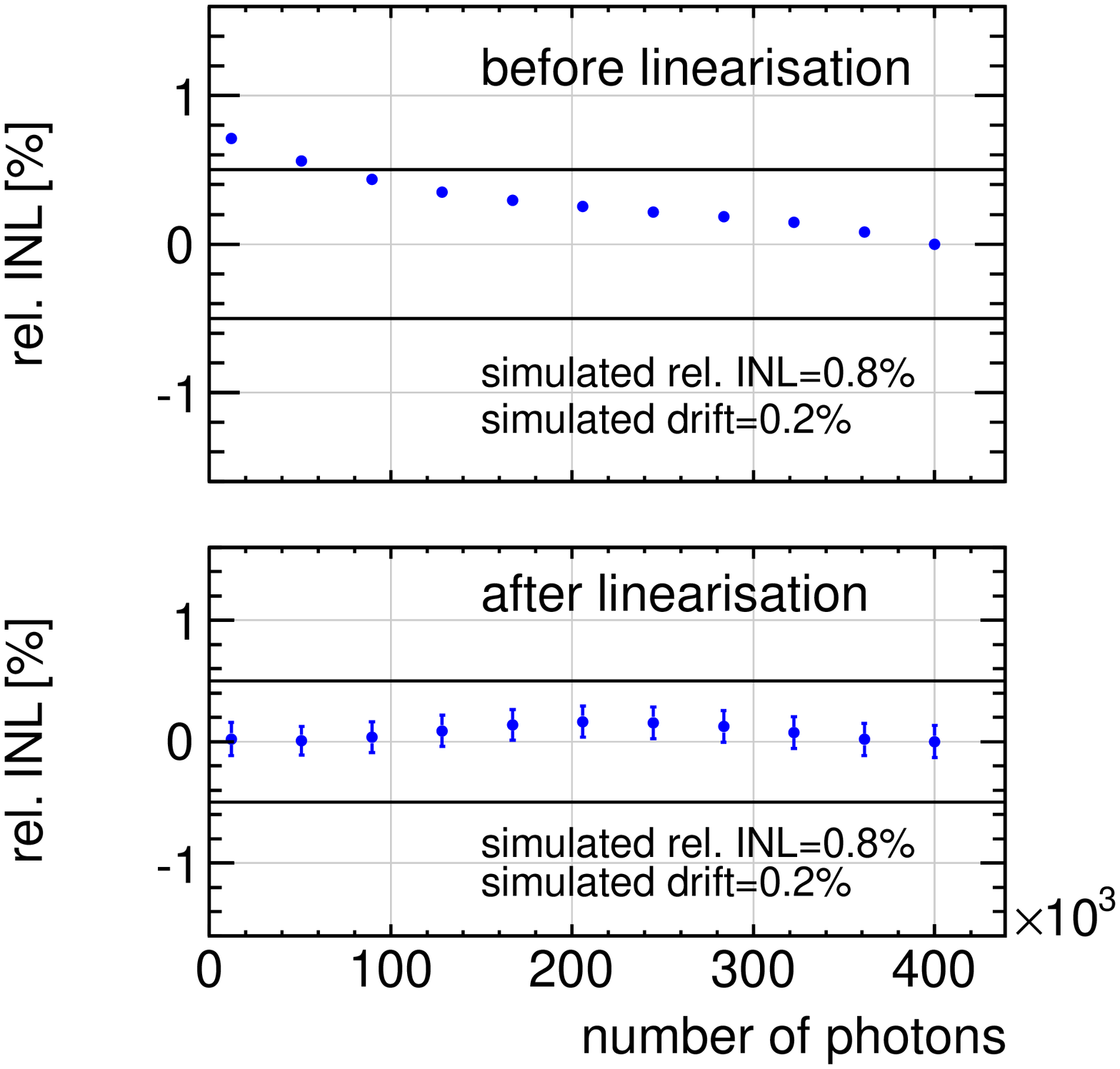}
%   \caption{$0.2\%$ drift of differential pulse during the scan of the dynamic range.}
  \label{fig:dnl:optimisation:drift:example02}
 \end{subfigure}
  \begin{subfigure}[b]{.49\linewidth}
  \centering
 \includegraphics[width=0.9\textwidth]{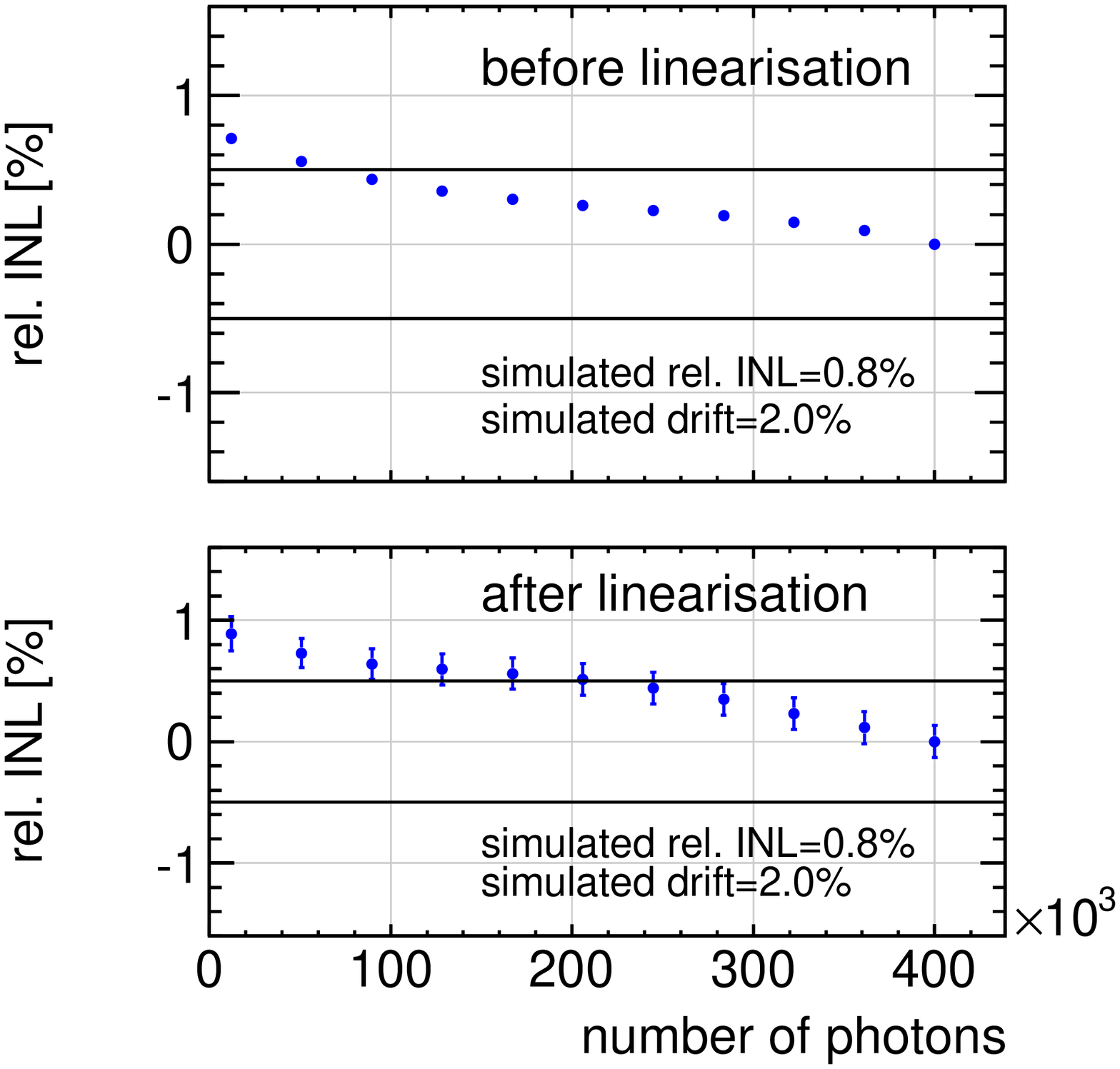}
% \includegraphics[width=0.9\textwidth]{figures/dnlmethod/optimisation/drift_nl008_extr0.eps}
%   \caption{$2.0\%$ drift of differential pulse during the scan of the dynamic range.}
  \label{fig:dnl:optimisation:drift:example20}
 \end{subfigure}
  \caption{Simulation of the effect of $0.2\%$ and $2.0\%$ drift in the light 
  intensity of the differential pulse during the scan of the dynamic range. The 
  horizontal line indicates the design goal of a nonlinearity smaller than 
  $0.5\%$. For a too large drift, the method does not sufficiently linearise 
  the detector response anymore.}
 \label{fig:dnl:optimisation:drift}
\end{figure}

\begin{figure}[htb]
 \centering
 \includegraphics[width=0.48\textwidth]{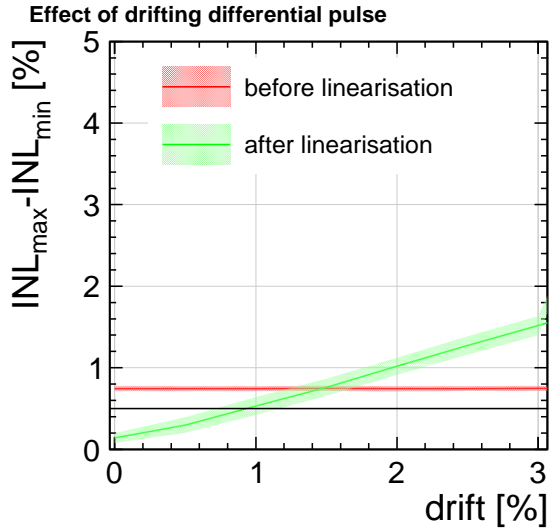}
 \caption{Effective nonlinearity in dependence of different simulated drifts of 
 the differential pulse. The drawn error band shows the RMS around the mean 
 value from all 200 simulated shapes of nonlinearities. In order to keep the 
 resulting nonlinearity below $0.5\%$, the differential pulse must not drift 
 by more than $0.5\%$.}
 \label{fig:dnl:optimisation:drift:result}
\end{figure}

One key requirement is the stability of the differential light pulse.
Figure~\ref{fig:dnl:optimisation:drift} shows the impact of a drift of the 
intensity of the differential pulse by $0.2\%$ and $2.0\%$ within 
the scan range on the resulting nonlinearity before and after the linearisation.
This figure illustrates that for small drifts the detector 
response can still be well linearised. However, once the drift gets too large 
the linearisation method does not work satisfactory anymore.
%Different measurement parameters, like the data point statistics and a drift 
%of the differential pulse have been simulated.
In order to estimate the maximal tolerable drift of the differential pulse, 
this parameter has been scanned in the simulation. Thereby, for each scan step,
200 randomly generated nonlinear transfer functions featuring the same total 
integrated nonlinearity of $0.8\%$ but different shapes have been simulated.
Figure \ref{fig:dnl:optimisation:drift:result} shows the average effective 
nonlinearity, defined as $\mathrm{INL}_\mathrm{max}-\mathrm{INL}_\mathrm{min}$ 
in the studied range, after the linearisation. The uncertainty band indicates 
the RMS of the resulting distribution around the central value. From the figure 
it gets clear that the differential pulse must not drift by more than $0.5\%$ 
during the scan of the dynamic range in order to still fulfil the goal of a 
linearisation better than $0.5\%$.

\begin{figure}[htb]
 \centering
 \includegraphics[width=0.49\textwidth]{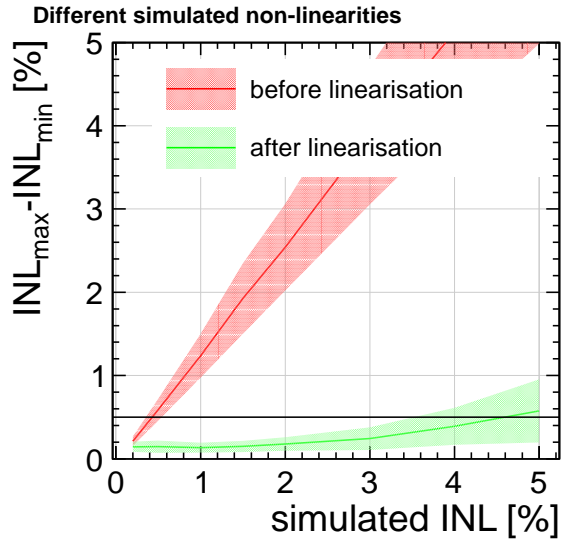}
  \caption{Effective nonlinearity in dependence of the simulated relative INL. 
  The drawn error band shows the RMS around the mean value from all 200 
  simulated shapes of nonlinearities. The proposed linearisation method still 
  works for a relative INL of up to $3.5\%$.}
 \label{fig:dnl:optimisation:nl}
\end{figure}

The presented linearisation method is based on the assumption of only small 
detector nonlinearities (cf. Equation \eqref{eq:smallnlapprox}). Therefore, 
also the limitations of the method with respect to the underlying 
nonlinearity needs to be studied. Figure \ref{fig:dnl:optimisation:nl} shows 
the effective nonlinearity before and after linearisation for different 
simulated nonlinearities. One can see that nonlinearities of up to $3.5\%$ can 
still be corrected successfully with the proposed approach such that the 
effective nonlinearity stays below $0.5\%$. This range is sufficient for 
photomultipliers, which are intrinsically very linear devices.

%*******************************************************************************
\subsection{Requirements for the calibration system}
\label{subsec:req}
%*******************************************************************************

The differential calibration approach has the advantage that the requirements 
for the calibration light source is reduced. However, it still needs to fulfil 
a few specifications:
\begin{description}
 \item[General design.]
   The calibration system must consist of two independent light sources, which can be 
   independently adjusted in their light intensity.
 \item[Stability of differential pulse.]
   The light pulse of the differential pulse defines the measure of the 
   derivative and, therefore, it must be very stable. In the previous section
   it has been shown that a relative drift of the differential pulse of 
   $0.5\%$ is tolerable in order to still meet the calibration requirements.
 \item[Dynamic range of base signal.]
  The light source providing the base pulses has to be adjustable in its light 
  intensity over the whole dynamic range of the planned Compton polarimeters 
  in order to be able to probe the differential nonlinearity within this range. 
  As can be seen from figure~\ref{fig:gas:expectedcomptone}, up to $200$
  Compton electrons per bunch crossing can be expected in a single detector channel,
  assuming the nominal instantaneous Compton luminosity.
%   A second light source is needed to provide the base light pulse.
%   Thereby it is important that the light source is able to probe the dynamic 
%   range that is relevant for the polarisation measurement.
%   Neither an absolutely calibrated nor a linearly behaving input signal is 
%   needed, since only the detector response enters in the correction algorithm.
\end{description}

Additionally, there are some more requirements, which are important in order 
to meet the conditions of the actual polarisation measurement at the ILC.
\begin{description}
 \item[Wavelength.]
  The light sensors at the Compton polarimeters are supposed to detect 
  Cherenkov light, whose wavelength intensity distribution peaks in the UV 
  range. Therefore, the calibration light source should emit light in this 
  wavelength range as well.
  %more details? 
  %plot with Cherenkov spectrum?
 \item[Short pulses.]
%   The light pulses have to be as short as possible.
  An ILC bunch with a longitudinal bunch length of $300\mum$ \cite{tdr} 
  traversing the Compton polarimeters creates a Cherenkov light pulse of 
  only $t=1\fs$. This is much shorter than the usual transit time of the 
  electron avalanche in a photomultiplier. Therefore, the light pulse of the 
  calibration source should be as short as possible, but at least as short as 
  the photomultiplier transit time, which is of the order of 
  $10\ns$ for photomultipliers we consider in out setups.% \cite{HamamatsuPMTbook}.
 \item[Applicable in detector design.]
  The light source should ideally be very small such that it can be integrated 
  in the existing polarimeter design.
%   According to the recent design 
\end{description}

\subsection{LED driver}
\label{subsec:leddriver}
\begin{figure}[htb]
 \begin{subfigure}[b]{.7\linewidth}
  \centering
  \includegraphics[width=0.97\textwidth]{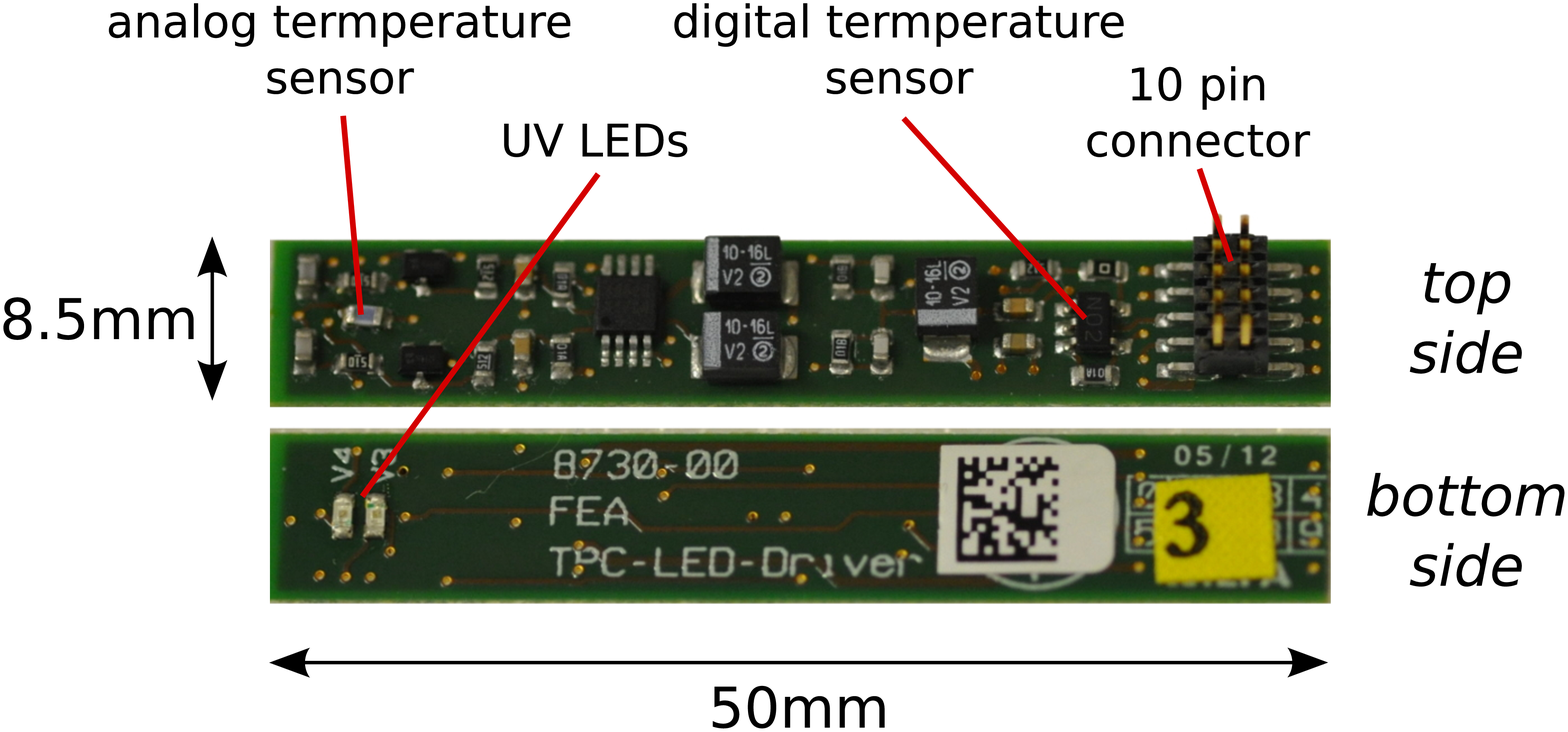}
  \caption{}
  \label{fig:gas:leddriver:driver}
 \end{subfigure}
 \begin{subfigure}[b]{.28\linewidth}
  \centering
  \includegraphics[width=0.97\textwidth]{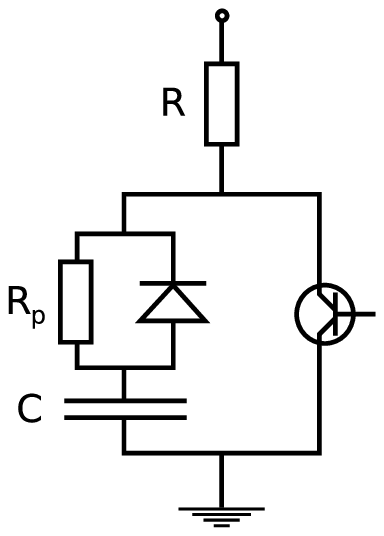}
  \caption{}
  \label{fig:gas:leddriver:schematics}
 \end{subfigure}
 \caption{(\subref{fig:gas:leddriver:driver}) Picture of the LED driver. The design is chosen such that it is applicable in the gas Cherenkov detector design. (\subref{fig:gas:leddriver:schematics}) LED driver circuit.}
 \label{fig:gas:leddriver}
\end{figure}

According to these specifications an LED driver has been developed based on the 
calibration light source of the CALICE tile hadron 
calorimeter \cite{Reinecke2011}. Figure~\ref{fig:gas:leddriver:driver} shows a
picture of the LED driver. It is equipped with two UV-LEDs which feature a peak
intensity at $\lambda=395\nm$ and a rather large spectral width of a 
few $10\nm$. The LED light intensity is defined by two external reference
voltages $U_\mathrm{LED1,2}$. Both LEDs receive the same trigger signal.

It is eye-caching that the dimensions of the board are rather small
($50\times8.5\mm^2$) such that the driver can easily be integrated in the 
existing Cherenkov detector design (see section~\ref{subsec:gas}).
Both UV-LEDs are located on the bottom side of the board, whereas the remaining 
electronic components are placed on top.
% In a final polarimeter Cherenkov detector assembly, one LED board is foreseen for each of the staggered Cherenkov detector channels, which sets the constraint of the tolerable width of the board.
As LEDs are rather temperature dependent devices, two temperature sensors are 
located on the board. An analogue temperature sensitive resistor 
(\texttt{Pt1000}) is placed directly opposite the LEDs on the top side. A 
second, digital temperature sensor (\texttt{DS18B20U}) is present at the other 
end of the board, which provides a 12-bit digitised temperature, read out via
a 1-wire bus. The board is connected via a 10-pin connector.

A simplified schematic of the LED driver circuit is shown in
Figure~\ref{fig:gas:leddriver:schematics}. This design \cite{Reinecke2011} allows
for very short light pulses. There are three phases in the generation of a 
light pulse:
Firstly, the capacitor $C$ gets charged via the resistors $R$ and $R_P$ ($R\gg R_P$)
driven by the external voltage $U_\mathrm{LED}$.
Secondly, when the trigger arrives, the transistor is closed and the 
capacitor discharges via $R_P$ and $R_\mathrm{LED}$. The LED is emitting light.
Thirdly, the resistance of the LED rises as the voltage at the  capacitor
drops due to the nonlinear $I-V$ behaviour of the LED. The discharging continues 
mainly via the parallel resistor $R_P$ and, thus, the LED light pulse is
quenched.

\begin{figure}[htb]
%  \begin{subfigure}[b]{.65\linewidth}
  \centering
  \includegraphics[width=0.8\textwidth]{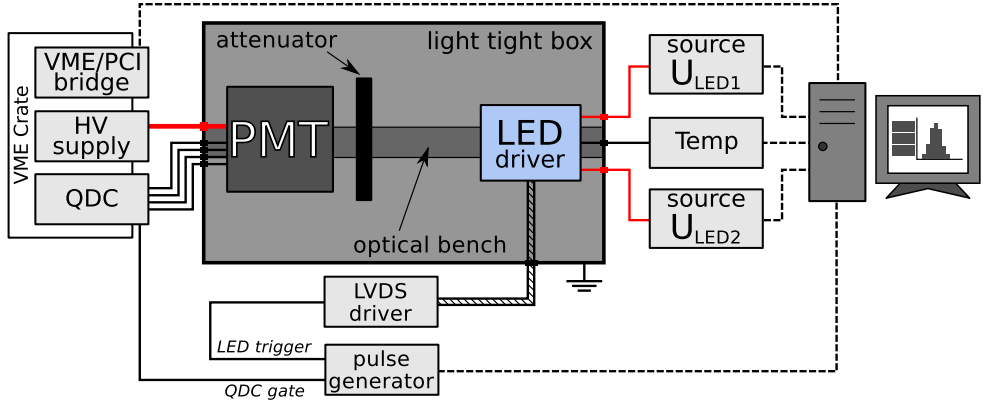}
  \caption{Schematic picture of test setup for characterising the LED driver board and for measuring the differential nonlinearity of a photomultiplier.}
   \label{fig:gas:setup:schematics}
%  \end{subfigure}
%  \begin{subfigure}[b]{.35\linewidth}
%   \centering
%   \includegraphics[width=0.97\textwidth]{figures/dnlmeasurement/setup/setup_real2.eps}
%   \caption{}
%   \label{fig:gas:setup:photo}
%  \end{subfigure}
%  \caption{(\subref{fig:gas:setup:schematics}) Schematic picture of test setup for characterising the LED driver board and for measuring the differential nonlinearity of a photomuiltiplier. (\subref{fig:gas:setup:photo}) Picture of the optical bench with LED board (left) and photomultiplier (right).}
%  \label{fig:gas:setup}
\end{figure}

The LED board has been carefully characterised in a dedicated test setup.
Figure~\ref{fig:gas:setup:schematics} shows a schematic overview of the setup.
It consists of a light tight box with an optical bench inside, on which the LED 
driver as well as a photomultiplier is mounted. 
%Figure~\ref{fig:gas:setup:photo}) shows the optical inside the light-tight box.
%The mounted LED driver is on the left and the photomultiplier right.
As photodetector a segmented photomultiplier by Hamamatsu 
(\texttt{R5900-03-M4}) has been selected. This type has also been operated 
in the gas Cherenkov detector prototype described in section~\ref{subsec:gas}.
For a coarse adjustment of the light intensity, different neutral density 
filters can be inserted into the optical path. In the standard setup, the 
photomultiplier is read out by a 12-bit QDC.

\begin{figure}[htb]
 \begin{subfigure}[b]{.5\linewidth}
  \centering
  \includegraphics[width=0.97\textwidth]{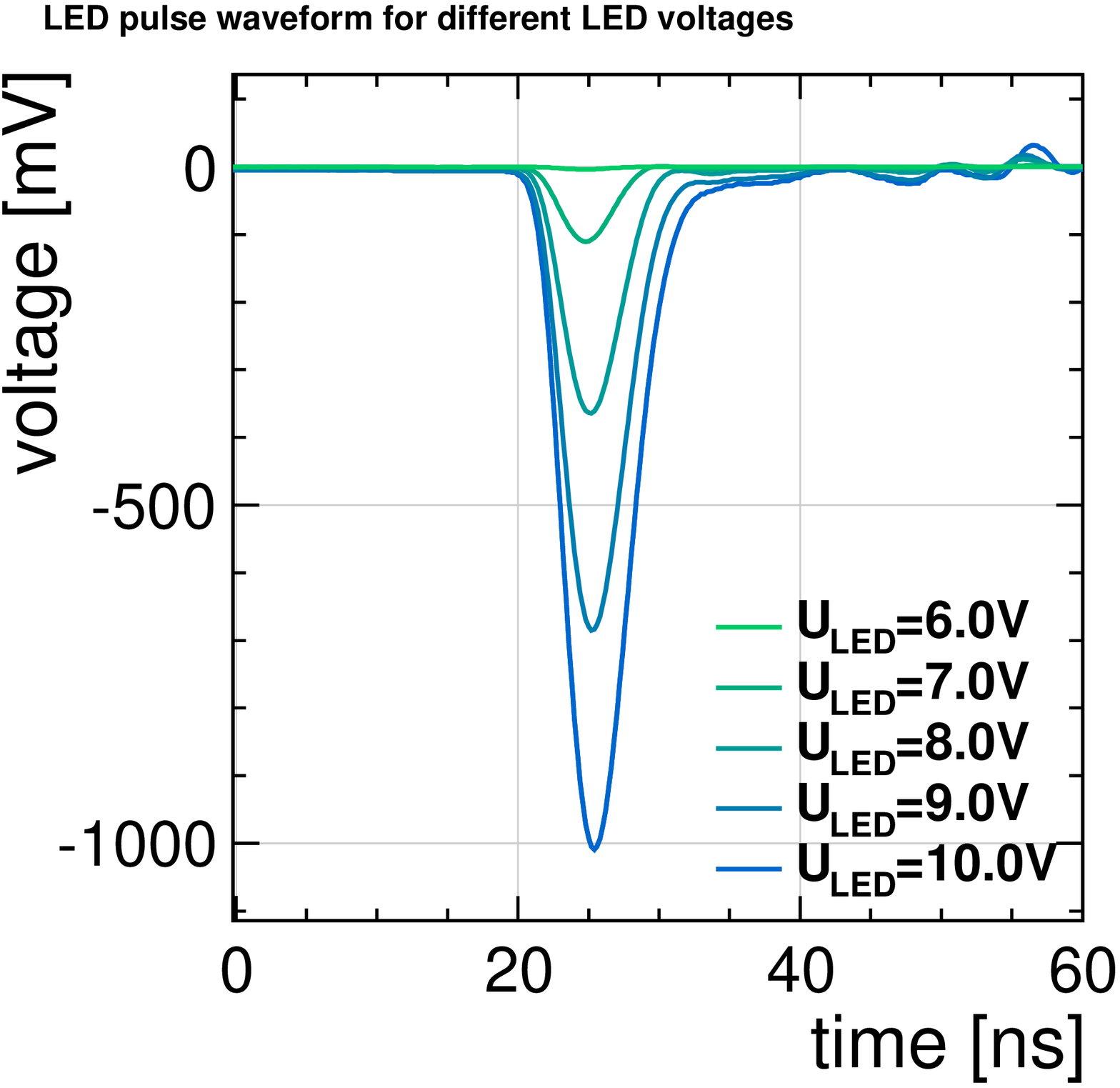}
  \caption{}
  \label{fig:gas:leddriver:pulselength}
 \end{subfigure}
 \begin{subfigure}[b]{.5\linewidth}
  \centering
  \includegraphics[width=0.97\textwidth]{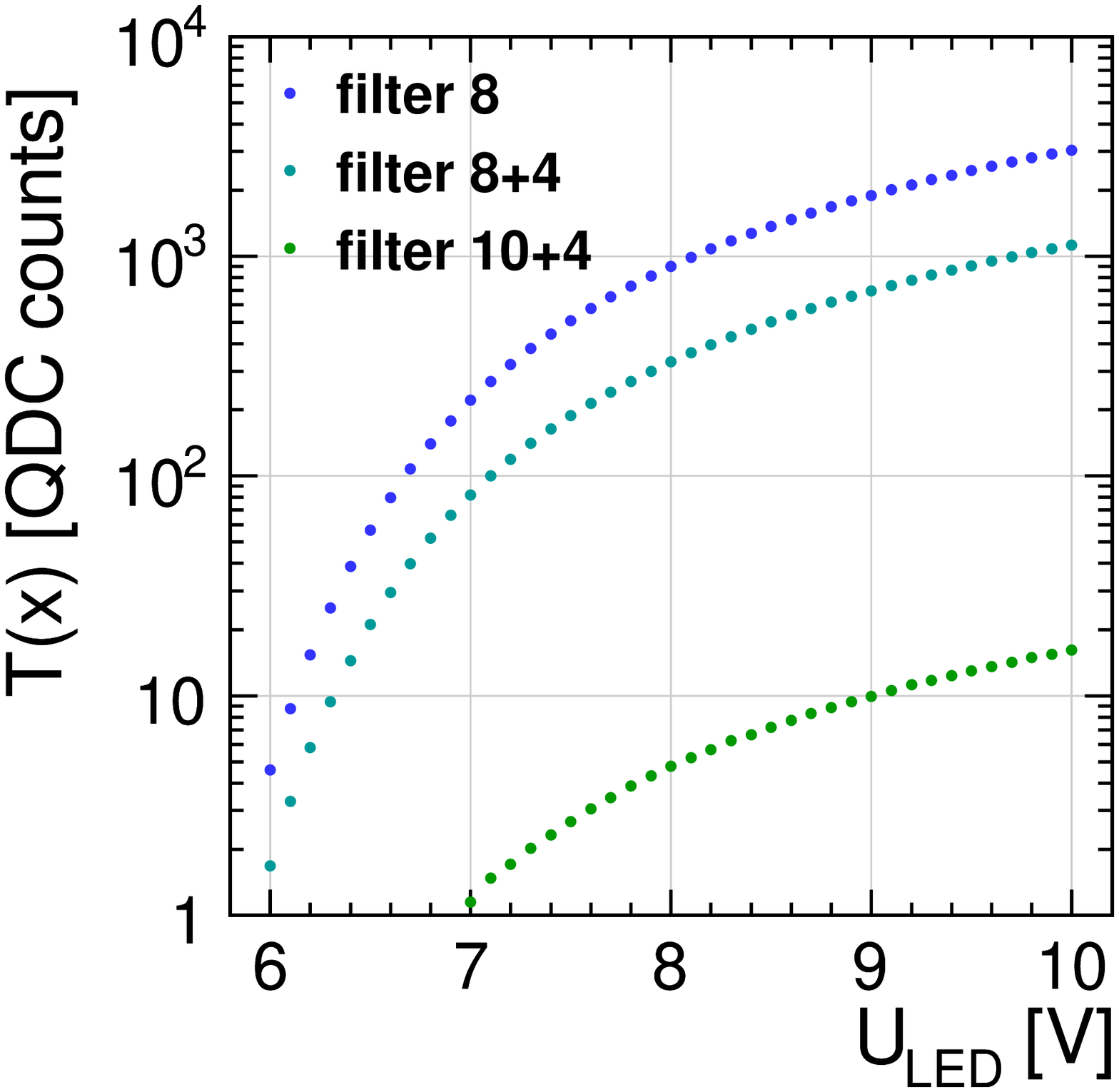}
  \caption{}
  \label{fig:gas:leddriver:dynamicrange}
 \end{subfigure}
 \caption{(\subref{fig:gas:leddriver:pulselength}) Recorded waveforms of the photomultiplier response for different reference voltages. The pulses are shorter than $10\ns$. (\subref{fig:gas:leddriver:dynamicrange}) Dynamic range of the LED driver for different filter configurations. The expected dynamic range of the ILC polarimeter can be covered.}
 \label{fig:gas:leddriver:characterisation}
\end{figure}

In order to resolve the pulse length of light pulses emitted by the LED driver, 
the photomultiplier has been connected to an oscilloscope at first. 
Figure~\ref{fig:gas:leddriver:pulselength} shows the recorded wave forms for 
different reference voltages $U_\mathrm{LED}$ averaged over 5120 individual 
light flashes. It is visible that the measurable pulse length is below $10\ns$ 
and, thus, the driver fulfils the described pulse length requirement. The 
observable wiggles at $t>40\ns$ can be attributed to photomultiplier after 
pulses and signal reflections because of imperfect signal termination at the 
oscilloscope. In order to exclude those nonlinear effects from the QDC 
measurements it is important to chose the gate carefully.

In Figure~\ref{fig:gas:leddriver:dynamicrange}, a QDC measurement of the 
dynamic range of the LED driver is depicted. From Monte-Carlo simulations it is
known that at the ILC polarimeters signals of up to 3000 QDC counts are 
expected in the most extreme detector channels
(c.f. Figure~\ref{fig:gas:expectedcomptone}). Thus, 
Figure~\ref{fig:gas:leddriver:dynamicrange} proofs that the expected dynamic 
range can completely be covered by the LED driver. Furthermore, the dynamic
range of the LED driver can be adjusted easily with different filter 
configurations if needed.
%2500 QDC counts, 200 c.e., g=3.5*10^5, 25fm/bin

\begin{figure}[htb]
%  \begin{subfigure}[b]{.32\linewidth}
%   \centering
%   \includegraphics[width=0.97\textwidth]{figures/dnlmeasurement/premeasurements/stabilityhyst700V10e5.eps}
%   \caption{}
%   \label{fig:gas:leddriver:ledvoltagejumps:overview}
%  \end{subfigure}
 \begin{subfigure}[b]{.5\linewidth}
  \centering
  \includegraphics[width=0.97\textwidth]{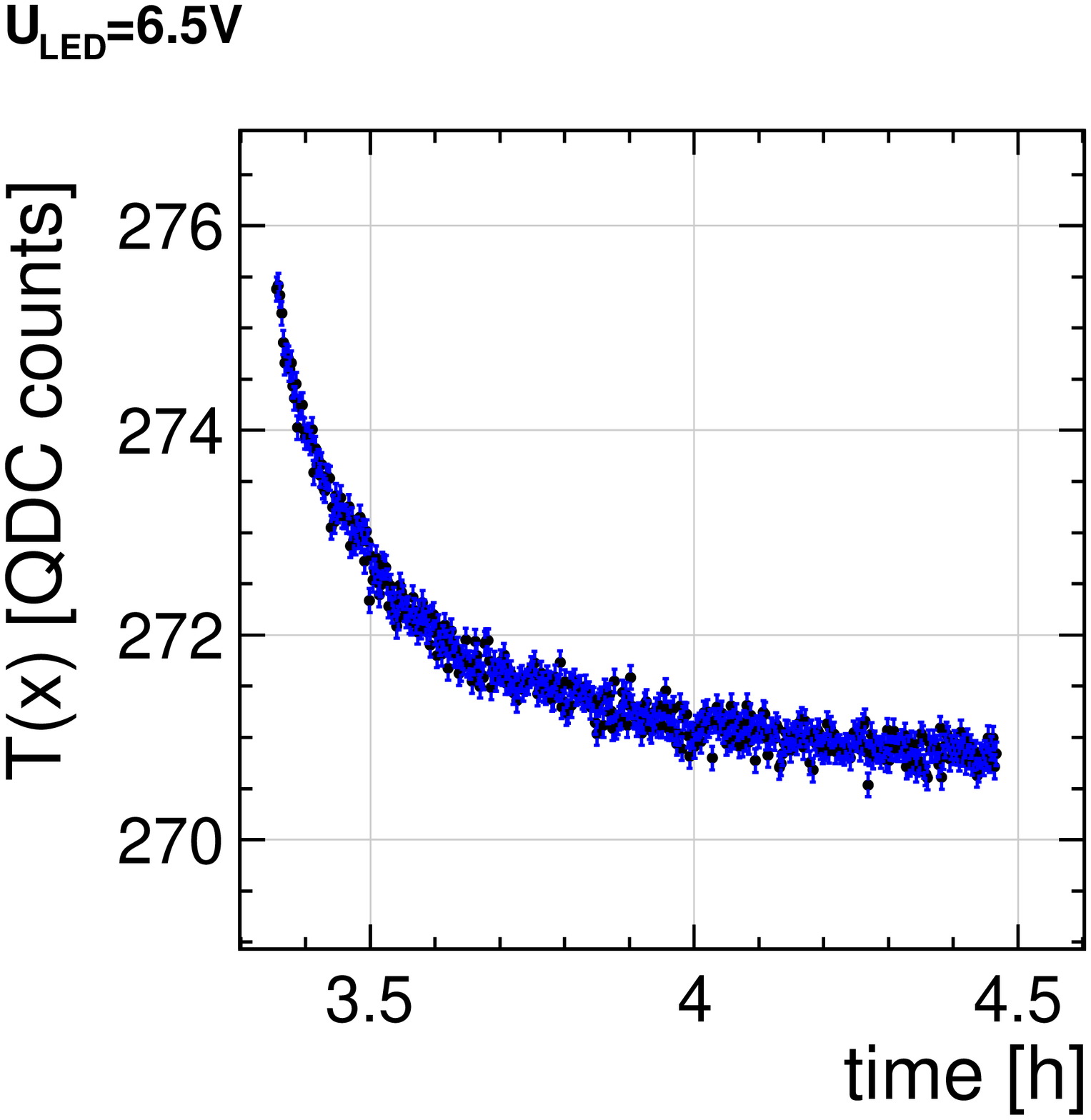}
  \caption{}
  \label{fig:dnlmeasurement:ledvoltagejumps:down}
 \end{subfigure}
 \begin{subfigure}[b]{.5\linewidth}
  \centering
  \includegraphics[width=0.97\textwidth]{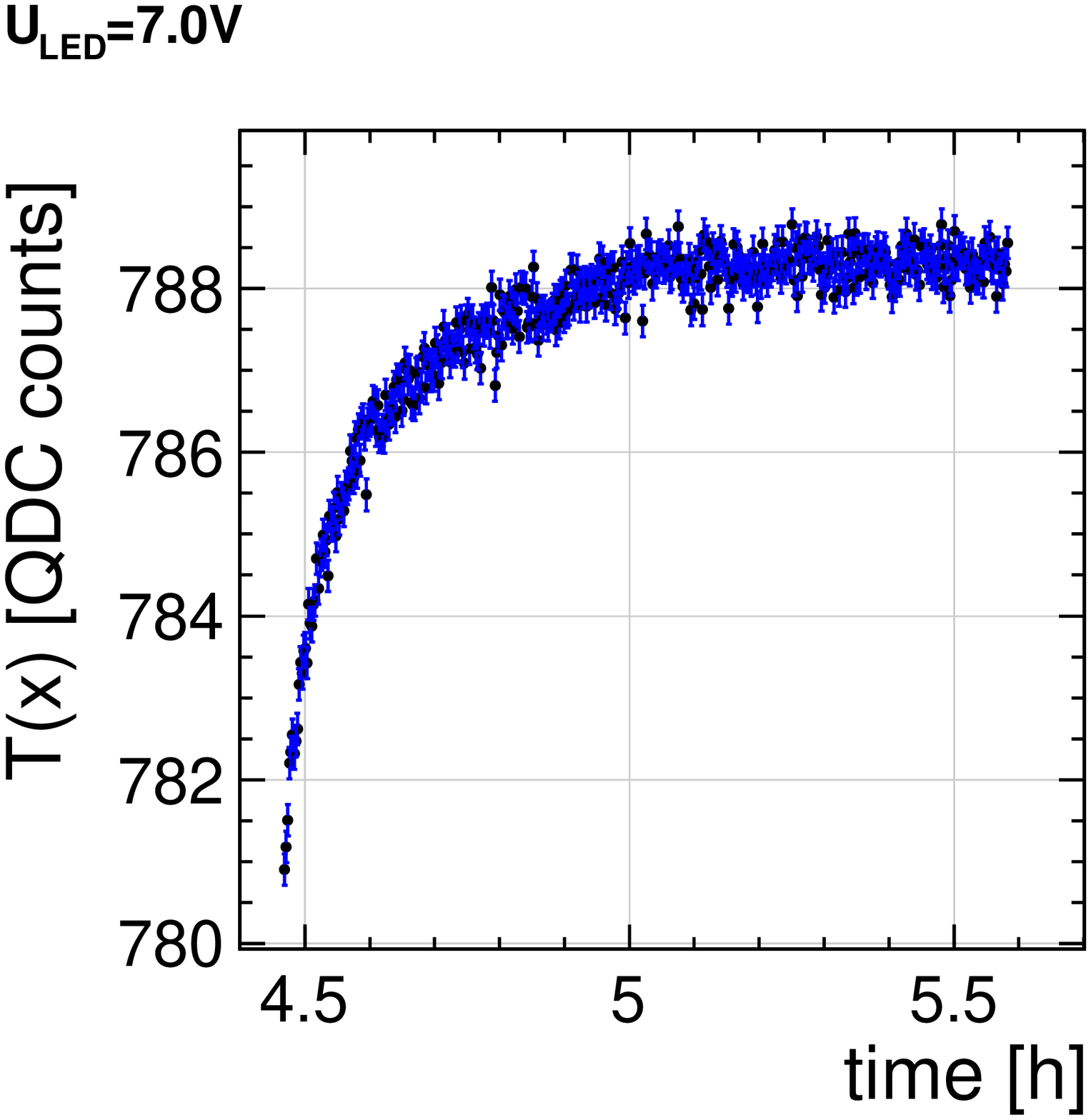}
  \caption{}
  \label{fig:dnlmeasurement:ledvoltagejumps:up}
 \end{subfigure}
%   \centering
%   \includegraphics[width=0.32\textwidth]{figures/dnlmeasurement/premeasurements/stabilityhyst700V10e5.eps}
% %   \includegraphics[width=0.32\textwidth]{figures/dnlmeasurement/premeasurements/stabilityhyst700V10e5zoom1.eps}
% %   \includegraphics[width=0.32\textwidth]{figures/dnlmeasurement/premeasurements/stabilityhyst700V10e5zoom2.eps}
% %   \includegraphics[width=0.32\textwidth]{figures/dnlmeasurement/premeasurements/stabilityhyst700V10e5zoom3.eps}
%   \includegraphics[width=0.32\textwidth]{figures/dnlmeasurement/premeasurements/stabilityhyst700V10e5zoom12.eps}
%   \includegraphics[width=0.32\textwidth]{figures/dnlmeasurement/premeasurements/stabilityhyst700V10e5zoom22.eps}
%   \includegraphics[width=0.32\textwidth]{figures/dnlmeasurement/premeasurements/stabilityhyst700V10e5zoom32.eps}
    \caption{Pedestal corrected mean position of the QDC spectra for changing LED voltages $U_\mathrm{LED}=6.5\V$ and $U_\mathrm{LED}=7.0\V$. About 40 minutes after a voltage change a stable mean position is reached.}
 \label{fig:dnlmeasurement:ledvoltagejumps}
\end{figure}

In order to test the stability of the base pulse, the PMT response to light 
pulses at different fixed LED reference voltages has been measured for about 
one hour each. Figure~\ref{fig:dnlmeasurement:ledvoltagejumps} displays two of 
these measurements at different voltage levels. The data points represent the 
pedestal corrected mean positions of a QDC spectra corresponding to $N=10^5$ 
light flashes, thus $T(x)$ in the terminology introduced in section~\ref{subsec:dnl}. 
In Figure~\ref{fig:dnlmeasurement:ledvoltagejumps:down}, the 
reference voltage has been set from a higher level to $U_\mathrm{LED}=6.0\V$.
In contrary, in Figure~\ref{fig:dnlmeasurement:ledvoltagejumps:up} the voltage 
level has been raised from a lower level to $U_\mathrm{LED}=6.5\V$. A dragging 
behaviour in opposite directions can be observed in the figures although the 
reference voltage was constant. After about $40$ minutes the light intensity 
stabilises. This is important to note, since for the proposed nonlinearity 
measurement the base pulse is altered every few minutes. In order to achieve 
reliable results for the nonlinearity measurement, the base pulse should always 
be scanned in both directions such that the dragging effect can be cancelled by 
averaging.

The differential pulse stability has been checked on the absolute scale for 
12~hours, which corresponds roughly to the time of a full nonlinearity 
measurement run in the shown setup. A drift of at maximum $0.41\%$ has been 
observed, which is within the tolerable range according to Monte-Carlo 
simulations (see Section~\ref{subsec:dnl:mc}).

% \begin{figure}
% %  \begin{subfigure}{.32\linewidth}
%   \centering
%   \includegraphics[width=0.5\textwidth]{figures/dnlmeasurement/premeasurements/stability7V6V700V10e5.eps}
% %   \caption{}
% %   \label{fig:dnlmeasurement:pulsestability7V6V:10e5}
% %  \end{subfigure}
% %  \begin{subfigure}{.32\linewidth}
% %   \centering
% %   \includegraphics[width=\textwidth]{figures/dnlmeasurement/premeasurements/stabilitypull7V6V700V10e5.eps}
% %   \caption{}
% %   \label{fig:dnlmeasurement:pulsestability7V6V:pull}
% %  \end{subfigure}
% %   \begin{subfigure}{.32\linewidth}
% %   \centering
% %   \includegraphics[width=0.5\textwidth]{figures/dnlmeasurement/premeasurements/stability7V6V700V10e7.eps}
% %   \caption{}
% %   \label{fig:dnlmeasurement:pulsestability7V6V:10e7}
% %  \end{subfigure}
%  \caption{Stability of the differential light pulse of LED~2. $\Us{LED1}=7\V$ and  $\Us{LED2}=6\V$ are chosen. $\Dmean$ is determined from $N=10^5$ light pulses.}
% %  \caption{Stability of the differential light pulse of LED~2. $\Us{LED1}=7\V$ and  $\Us{LED2}=6\V$ are chosen. (\subref{fig:dnlmeasurement:pulsestability7V6V:10e5}) $\Dmean$ determined from raw data of $N=10^5$ light pulses. (\subref{fig:dnlmeasurement:pulsestability7V6V:pull}) Corresponding pull distribution. (\subref{fig:dnlmeasurement:pulsestability7V6V:10e7}) Averaged $\Dmean$ over 100 data points.}
%  \label{fig:dnlmeasurement:pulsestability7V6V}
% \end{figure}

During the tests it has been observed that the differential pulse intensity 
depends on the base pulse voltage level. In order to avoid this cross talk for 
the following nonlinearity measurements, two independent LED driver boards have 
been used. In this modified setup, only one LED per LED driver board has been 
used. The other LED on the boards has been set to $U_\mathrm{LED}=0\V$.

Except from the cross-talk problem, the LED driver has proven to fulfil the 
requirements for the usage in the proposed differential nonlinearity 
measurement. In a future design, the two LEDs on one board will be operated in
independent circuits.

% 
% % =============================================================================
%###############################################################################
\section{Measurement Results and Application to the ILC}
\label{sec:meas_and_ILC}
%###############################################################################

In this section, we present a measurement of the nonlinearity for
one of the PMTs used in the prototype detector, and we discuss the implications
for ILC polarimetry. 

%*******************************************************************************
\subsection{Nonlinearity measurements}
\label{subsec:nlmeas}
%*******************************************************************************

\begin{figure}
\centering
 \begin{subfigure}{.48\linewidth}
  \centering
  \includegraphics[width=\textwidth]{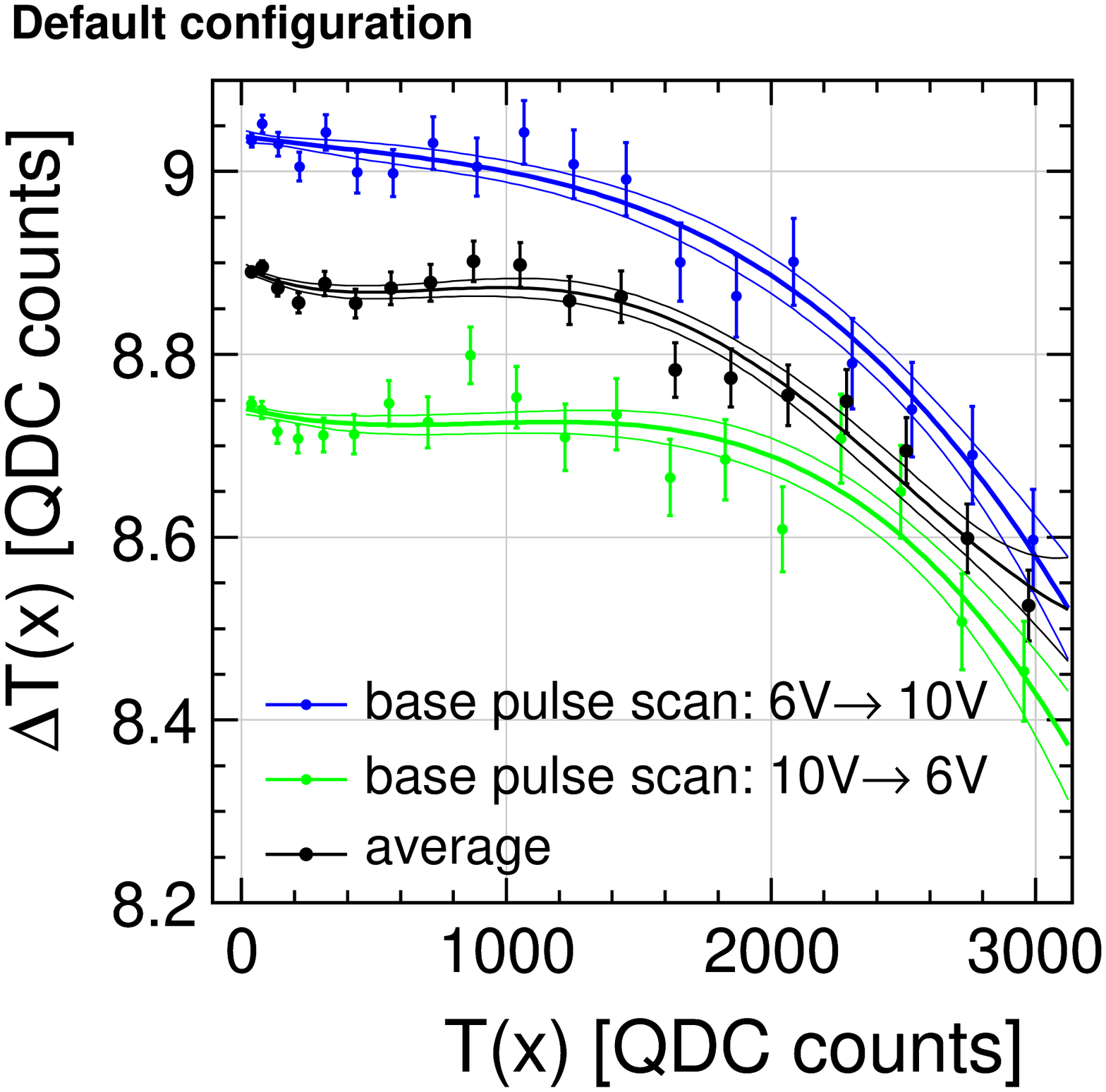}
  \caption{}
  \label{fig:dnlmeasurement:dnl:default:delta}
 \end{subfigure}
 \begin{subfigure}{.48\linewidth}
  \centering
  \includegraphics[width=\textwidth]{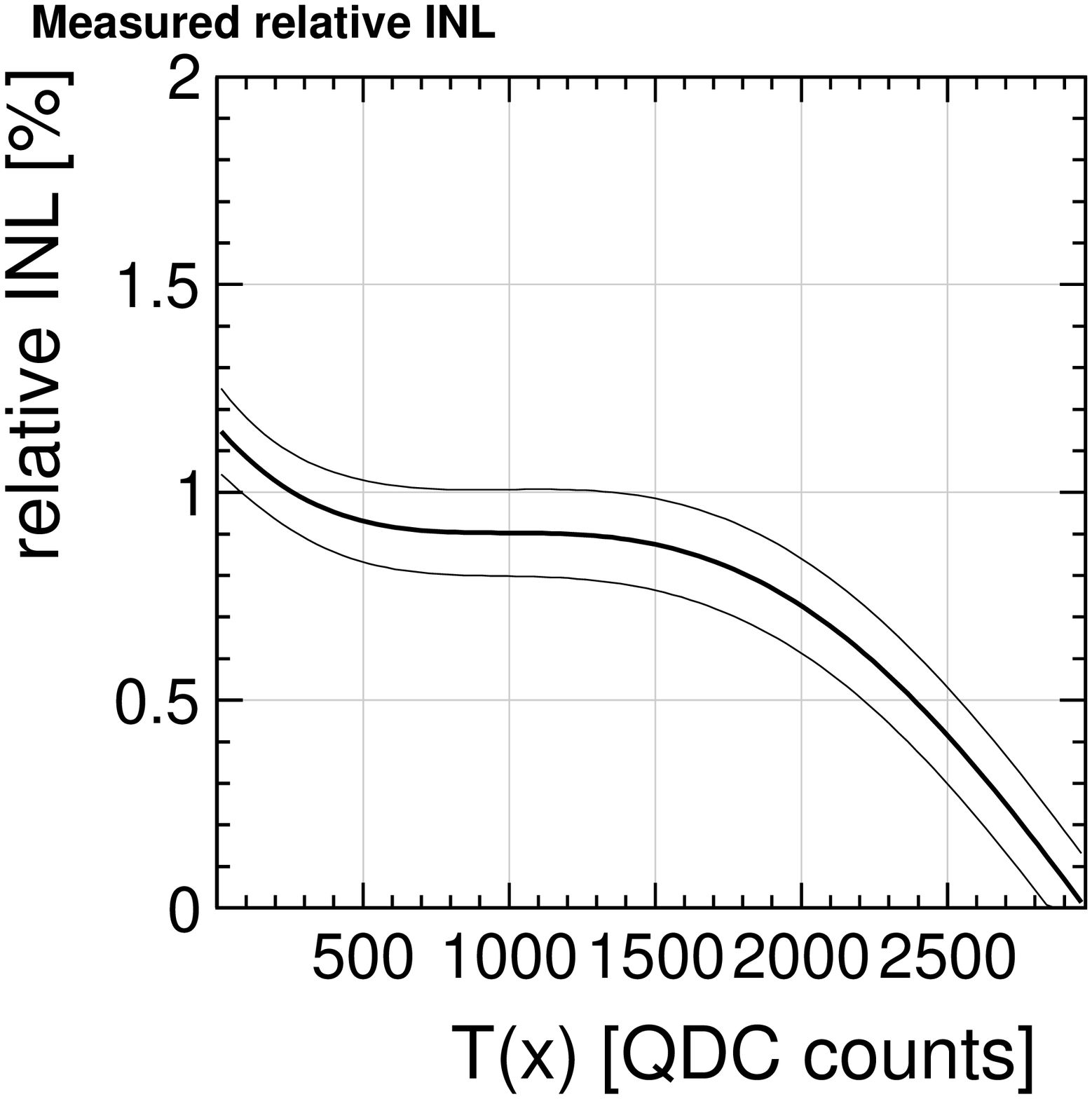}
  \caption{}
  \label{fig:dnlmeasurement:dnl:default:nl}
 \end{subfigure}

%  \begin{subfigure}{.48\linewidth}
%   \centering
%   \includegraphics[width=\textwidth]{figures/dnlmeasurement/dnl/meas_nl_abstract.eps}
%   \caption{}
%   \label{fig:dnlmeasurement:dnl:default:reconl}
%  \end{subfigure}
 \caption{(\subref{fig:dnlmeasurement:dnl:default:delta}) Measured $\Delta \T(x)$ in the default configuration for different scan directions of the base pulse. (\subref{fig:dnlmeasurement:dnl:default:nl}) Determined relative INL. 
  The absolute scale is not accessible with the presented method. }
 \label{fig:dnlmeasurement:dnl:default:}
\end{figure}

With the setup described in section~\ref{subsec:leddriver} we determined the nonlinearity
of one of the candidate PMTs to be employed in the Cherenkov detectors of the ILC polarimeters.
Figure~\ref{fig:dnlmeasurement:dnl:default:delta} shows the result of a differential nonlinearity measurement
of the photomultiplier \texttt{R5900-03-M4} operated at $700\V$. For each voltage setting of the base LED, responses $\T(x)$ to the base pulse alone and $\T(x + \Delta x)$ to base and differential pulse together
are determined from $10^5$ samples each. 

As already discussed, a slightly different $\Delta\T(x) = \T(x) - \T(x + \Delta x)$ is obtained depending on the scan direction of the base pulse.
The black data points show the average of both scan directions, which is used to determine the relative nonlinearity in Figure~\ref{fig:dnlmeasurement:dnl:default:nl}.
Since the nonlinearity is defined with respect to the detector response at the end of the dynamic range, the nonlinearity vanishes at this point by construction.
%Figure~\ref{fig:dnlmeasurement:dnl:default:reconl} visualises the measured detector response.
%In order to make the small nonlinearity visible, the nonlinearity has been multiplied by a factor of $10$ in this figure.
It can be observed qualitatively that the detector response starts to show a saturating behaviour, which is expected for photomultipliers at high light intensities.
The measurement has been repeated with different filter configurations and exchanged hardware and all measured nonlinearities are in agreement with the presented one~\cite{diss:vormwald}.

%(\subref{fig:dnlmeasurement:dnl:default:reconl})~Visualisation of the measured nonlinearity. The nonlinearity is magnified by a factor of ten in order to make the qualitative shape visible.

\begin{figure}
  \centering
%   \begin{subfigure}{.48\linewidth}
%   \centering
%   \includegraphics[width=\textwidth]{figures/dnlmeasurement/dnl/meas_default_corr.eps}
%   \caption{}
%   \label{fig:dnlmeasurement:dnl:default:corr}
%  \end{subfigure}
 \begin{subfigure}{.475\linewidth}
  \centering
  \includegraphics[width=\textwidth]{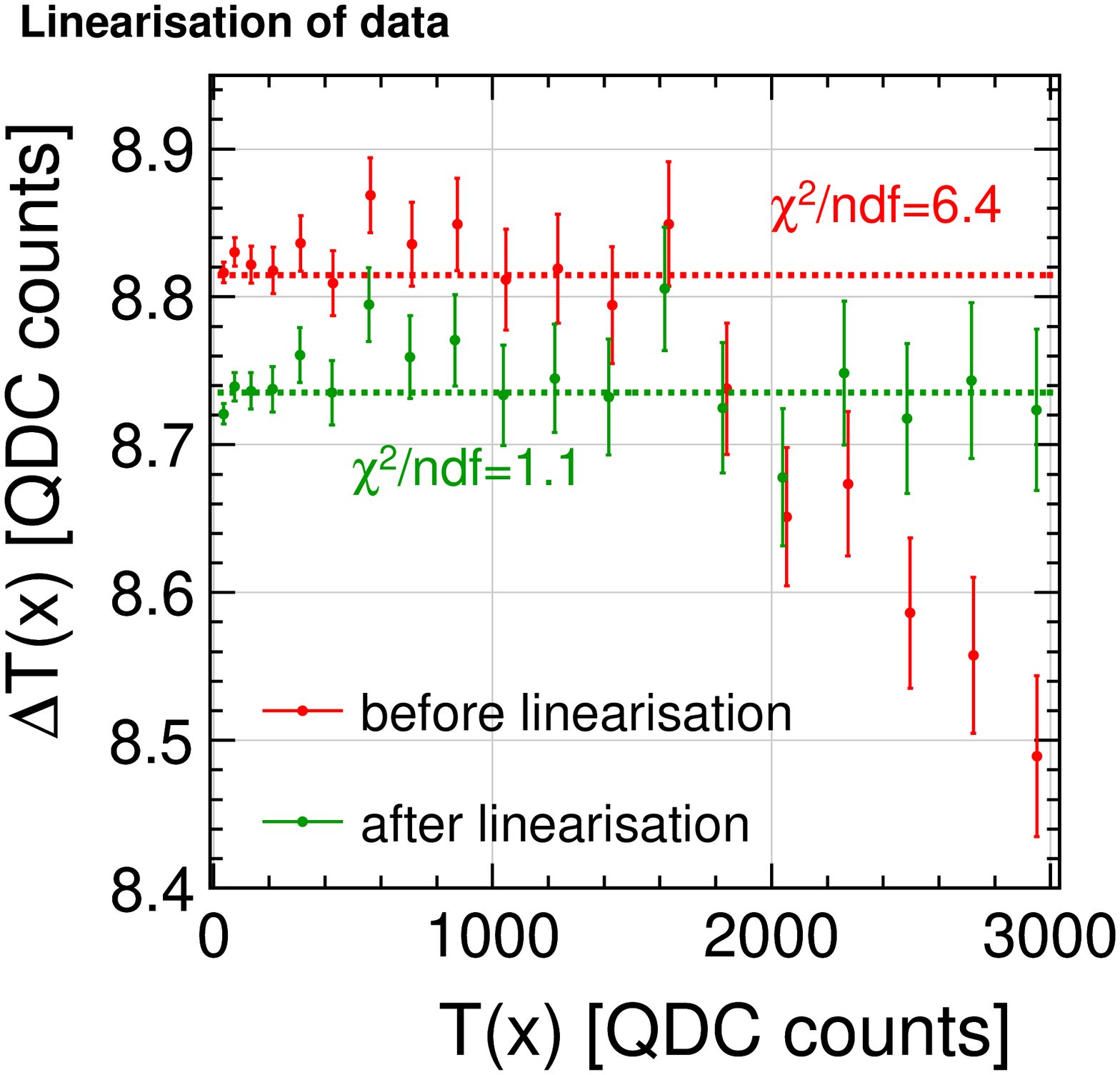}
  \caption{}
  \label{fig:dnlmeasurement:dnl:default:closuretest}
 \end{subfigure}
  \hspace{0.1cm}
 \begin{subfigure}{.475\linewidth}
  \centering
  \includegraphics[width=\textwidth]{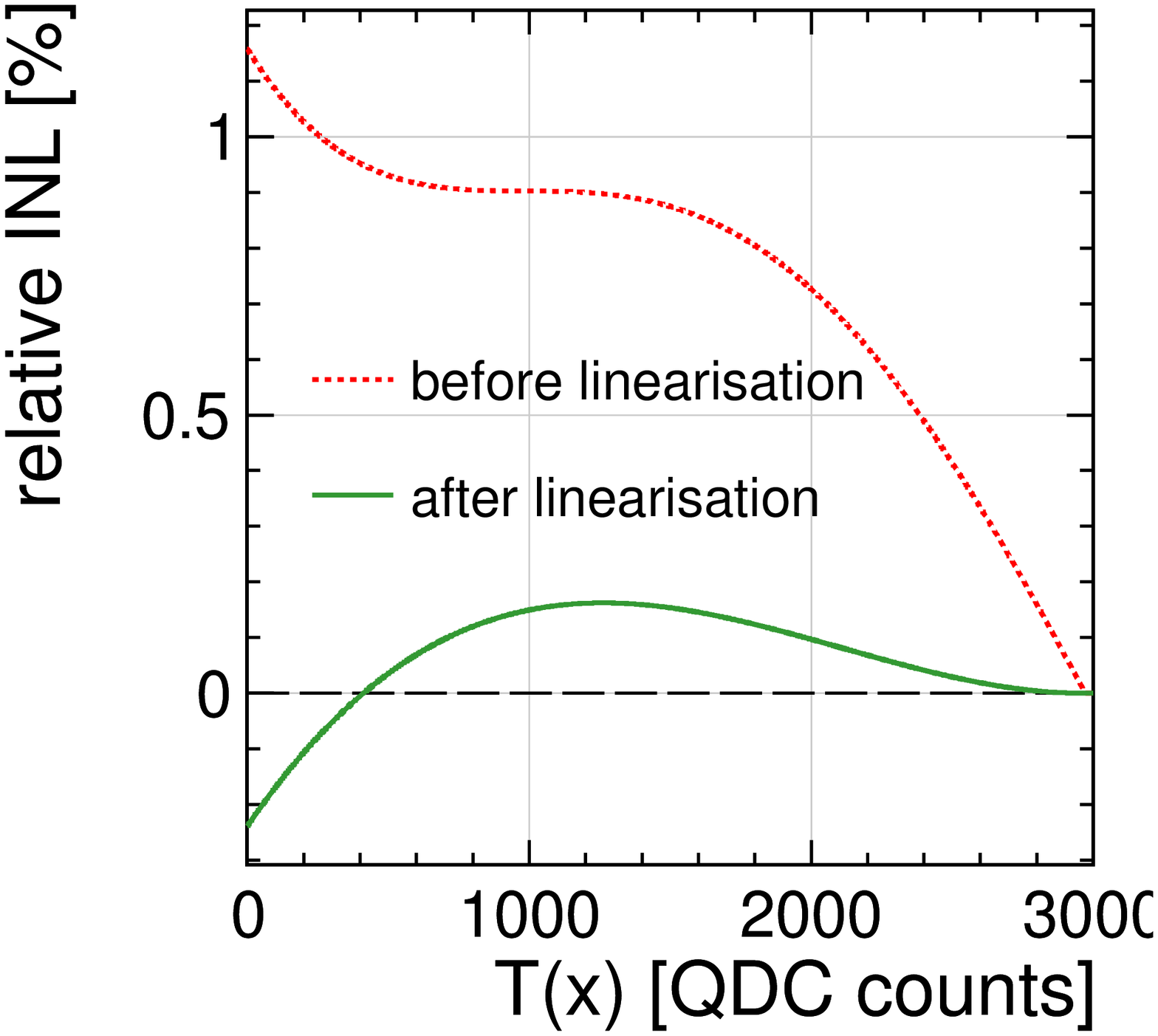}
  \caption{}
  \label{fig:dnlmeasurement:dnl:default:corrnl}
 \end{subfigure}
  \caption{
   (\subref{fig:dnlmeasurement:dnl:default:closuretest})~Independent data set before (red) and after (green) linearisation. (\subref{fig:dnlmeasurement:dnl:default:corrnl}) ~Visualisation of the measured nonlinearity from Figure~\ref{fig:dnlmeasurement:dnl:default:nl} and the residual nonlinearity after the linearisation. Linearity has been improved by about a factor of 5.}
 \label{fig:dnlmeasurement:dnl:closure}
\end{figure}

The measured nonlinearity can now be used for deriving a correction function. % as shown in Figure~\ref{fig:dnlmeasurement:dnl:default:corr}.
In order to demonstrate the correction power of this method, an independently measured data set has been linearised using the correction function.
Thereby, all base+differential and base pulse measurements have been individually corrected and from this corrected data $\Delta \T(x)$ has been calculated again.
Figure~\ref{fig:dnlmeasurement:dnl:default:closuretest} shows $\Delta \T(x)$ with respect to the base pulse mean position before and after the application of the linearisation algorithm.
It is clearly visible that after the linearisation the data points for $\Delta \T(x)$ are much more consistent with a constant than before.
Figure~\ref{fig:dnlmeasurement:dnl:default:corrnl} visualises the originally measured nonlinearity as well as the residual nonlinearity after correction, which is more than a factor $5$ smaller than before the correction.
%Figure~\ref{fig:dnlmeasurement:dnl:default:corrnl} visualises the nonlinearity of the corrected detector response. The residual nonlinearity is more than a factor $5$ smaller than before the correction.

  %(\subref{fig:dnlmeasurement:dnl:default:corr})~Correction function derived from Figure~\ref{fig:dnlmeasurement:dnl:default:nl}.

This demonstrates that the LED driver, the nonlinearity measurement and the linearisation method are
suitable to measure and correct PMT responses at the sub-percent level.

%*******************************************************************************
\subsection{Application to Polarimetry}
\label{subsec:pol}
\begin{figure}[hbt]
  \centering
% \begin{subfigure}{.475\linewidth}
%  \centering
  \includegraphics[width=0.5\textwidth]{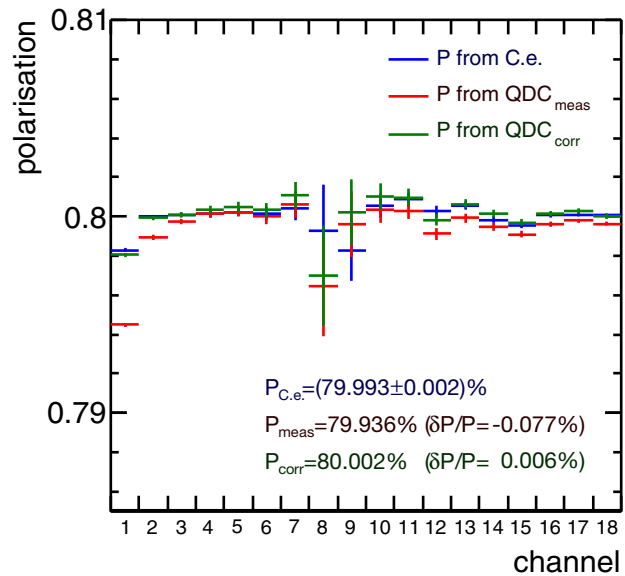}
%  \caption{}
%  \label{fig:dnlmeasurement:dnl:polmeas:all}
% \end{subfigure}
  \caption{Simulation of a polarisation measurement with the ILC upstream polarimeter based on
   the true number of Compton electrons per channel (blue, $P_{\mathrm{C.e.}}$), the mean values of the QDC spectra
    simulated with nonlinearity (red, $P_{\mathrm{meas}}$) and after applying the correction (green,    
    $P_{\mathrm{corr}}$).
  }
 \label{fig:dnlmeasurement:dnl:polmeas}
\end{figure}

%*******************************************************************************
As a final step, we consider the application to the ILC situation. For this purpose, the measured
nonlinearity from figure~\ref{fig:dnlmeasurement:dnl:default:nl} has been applied within the
fast Linear Collider Polarimeter Simulation \texttt{LCPolMC}~\cite{spintracking, Eyser2007}.
Figure~\ref{fig:dnlmeasurement:dnl:polmeas} shows the polarisation calculated in each detector
channel according to equation~\ref{eq:asymmetry}, based on three different estimators for
the number of Compton electrons $N$: The true number of Compton electrons (blue, $P_{\mathrm{C.e.}}$),
the mean values of the QDC spectra simulated with the measured nonlinearity from figure~\ref{fig:dnlmeasurement:dnl:default:nl} (red, $P_{\mathrm{meas}}$) and with the residual
nonlinearity after correction as shown in figure~\ref{fig:dnlmeasurement:dnl:default:corrnl} (green,    
$P_{\mathrm{corr}}$). In addition, the combined polarisation measurement obtained from the weighted
mean of the per-channel polarisations according to equation~\ref{eq:weights} is given in figure~\ref{fig:dnlmeasurement:dnl:polmeas}.

In this particular case, the measured nonlinearity causes a deviation of only $|\deltaP|=0.08\%$, which would still be within the envisaged uncertainty budget. Comparison with figure~\ref{fig:gas:status:nlrequirement} shows, however, that the nonlinearity measured for this specific
PMT has a rather benign shape, since for an INL of $1\%$ a mean deviation of $|\deltaP|=0.14\%$ is expected, with an RMS of $\pm 0.4\%$. 
%Although deviation of the average polarisation is small for the measured INL, the correction reduces the effect by more than an order of magnitude, thus reducing the contribution of the nonlinearity to the total error budget to a negligible level.
 
Considering the channel-dependency, the measured polarisation values are most obviously distorted in the channels with large absolute differences between the count rates for parallel and opposite laser and beam helicities, i.e.\ at small channel numbers. While at a first glance it could be argued that channels which measure a polarisation different from the average could simply be excluded, a closer look reveals that this is not sufficient: For all channels with a significant statistical weight, and especially those
above the zero crossing of the asymmetry (channel numbers larger than $10$), the uncorrected polarisation (red) is consistently smaller than the value expected from the number of Compton electrons (blue).
Furthermore, the dependence of the measured polarisation on the channel number is not a unique indicator for the presence or absence of nonlinearities, but can also be influenced by misalignments or tilts between detector and beamline~\cite{quartzdet}. Thus, the differential nonlinearity measurement is essential for a precise monitoring of the detector nonlinearity and its separation from other effects.

%When comparing to figure~\ref{fig:gas:expectedcomptone}, this corresponds to channels where the signal is below $1000$
%QDC counts for both laser helicities. In this low- and mid-intensity range, the photomultiplier response is already very linear, which manifests itself in a rather constant correction factor. However the effect of the nonlinearity and the improvement due to the calibration procedure remains clearly visible for all channels.  When not corrected for, the nonlinearity shifts the average polarisation by about $\deltaP=0.69\percent$, which is by far outside the uncertainty budget.

After applying the determined correction to the QDC mean values, the obtained channel-wise polarisation values $P_{\mathrm{corr}}$ (green) reproduce the reference values from the ``true'' numbers of Compton electrons nearly perfectly. This lays the foundation to disentangle effects due to
nonlinearities from others caused e.g.\ by the above-mentioned misalignments and tilts in a real polarimeter system. The relative deviation of the average polarisation from the Compton electron-based value becomes $0.006\percent$, thus more than a factor ten smaller than before, and a factor $16$ smaller the effect foreseen in the uncertainty budget in table~\ref{tab:polsys}. This shows that the linearisation based on our test setup would be fully sufficient for the ILC polarimeters, and even has the potential to reduce the impact of
nonlinearities to a negligible level.

%The dependence of the measured polarisation on the channel number as seen in the blue histogram in figure~\ref{fig:dnlmeasurement:dnl:polmeas} is not a unique indicator for the presence or absence of nonlinearities, but can also be influenced by misalignments or tilts between detector and beamline~\cite{quartzdet}. Thus, the differential nonlinearity measurement is essential for a precise monitoring of the detector nonlinearity and its separation from other effects.

\begin{figure}
  \centering
  \includegraphics[width=0.7\textwidth]{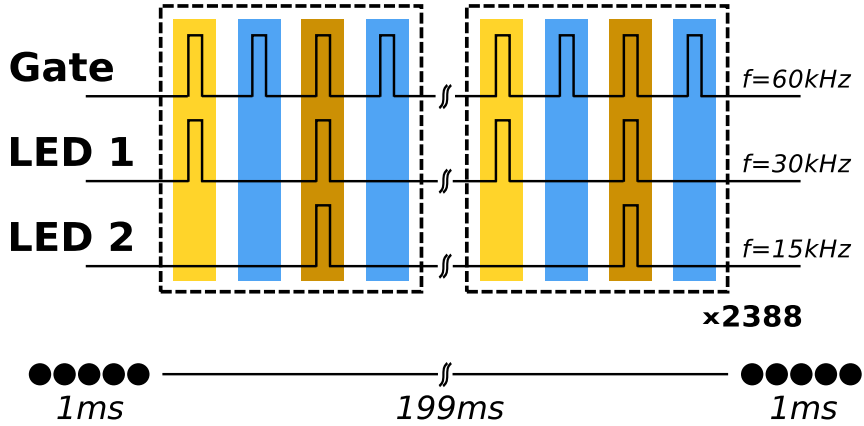}
 \caption{Possible trigger scheme for DNL measurements at the ILC. Between two bunch trains 2985 measurements of pedestal (blue), base light pulse (yellow), and base+differential (orange) pulse could be performed. This scheme allows for operating each LED with a constant frequency between the bunch trains. During a traversing bunch train both LEDs must not trigger.}
  \label{fig:dnlmeasurement:gateILC}
\end{figure}

%A careful characterisation of the photomultipliers before the actual operation in the ILC polarimeters is essential.
Beyond a careful initial characterisation of the photomultipliers in the lab before the actual installation in the ILC polarimeters, an in-situ monitoring of the detector calibration is clearly required to maintain the achieved
precision over many years of ILC operation. A possible measurement scheme at the ILC during a run is depicted in Figure~\ref{fig:dnlmeasurement:gateILC}. It makes use of the pulsed beam structure at the ILC in which each bunch train is separated in time by $199\ms$. The polarimeters are idle in this time such that this period can be utilised for calibration measurements.
Operating the LEDs at different frequencies $f_{\mathrm{LED1}}=2f_{\mathrm{LED2}}=30\kHz$ would allow for measuring $N=2388$ data points $(\T(x); \Delta \T(x))$ assuming that $80\%$ of the time between two trains can effectively be used.
In order to reach the demanded precision, $N=3\times10^6$ data points per base pulse voltage have to be acquired (c.f. Section~\ref{subsec:dnl:mc}), which is possible within less than $5$ minutes.
Thus, the whole dynamic range could be scanned back and forth in $10$ steps within less then $1.5$ hours in parallel to a normal ILC physics run.

In order to make this operation mode possible, small modifications of the LED driver board are necessary: Obviously, it must be synchronised with the ILC clock. Furthermore, 
both LEDs need to be able to be triggered independently, which can be easily realised in the current design.
This decoupling of the trigger circuit could also solve the observed cross talk between the two LEDs.

% 
% % =============================================================================
\section{Conclusions}
\label{sec:conclusion}
In this paper, we presented a calibration system for 
the Compton polarimeters at the ILC. The baseline
technology to measure the Compton scattered electrons
is a gas Cherenkov detector, read-out by single- or
multi-anode PMTs. One of the key performance parameters
is the linearity of such a detector. In order to keep
the corresponding contribution to the systematic uncertainty
of the polarisation measurement below $0.1\percent$, the  
nonlinearities must be controlled to $0.5\percent$ or better.
This is required over the whole dynamic range of the measurement,
which stretches over an order of magnitude in the number of 
Compton scattered electrons per detector channel. The 
most populated channel has an average count rate of $100-200$
electrons per bunch crossing, depending on the achieved
laser-electron luminosity.

At the level of $0.5\percent$, a calibration via an external 
absolute reference light source is not trivial. The problem
of calibrating the linearity of the light source can be circumvented
by measuring directly the differential nonlinearity with two short (few ns)
light pulses:
The first pulse must be tunable to cover the whole dynamic range, but
no special precision or linearity is required. The second pulse has a 
fixed size small compared to the maximal size of the base pulse. Again,
its exact size does not need to be known, but it  must be very stable,
with not more than $0.5\percent$ relative drift during a calibration cycle.
The differential nonlinearity can then be measured from the response
of the detector to the base pulse alone and to both pulses together.
In case of the application at the ILC polarimeters, about $10$ scan points for the base pulse with
$3\times10^6$ samples each are sufficient in order to achieve the required precision.

An LED driver providing these pulses has been designed and tested successfully
in a setup using one of the candidate PMTs to be employed in the gas Cherenkov
detectors of the ILC Compton polarimeters.
In this setup, a correction of the nonlinearities to the per-mille level could
be achieved. Transferring the result to a simulated polarisation measurement
shows a negligible impact of the nonlinearity on the polarisation measurement
after the correction. Finally, we propose a scheme how to collect calibration
data in-situ during the physics runs of the ILC, exploiting the rather long
gaps of $199$\,ms between the ILC bunch trains. A full calibration would then
be completed within $1.5$ hours, while more frequent updates of the calibration
can be obtained by a sliding window technique. 

In summary, the performance of the presented calibration system is sufficient to 
control the contribution of detector nonlinearities to the total uncertainty of the
polarisation measurement to much better than $0.1\percent$. This could reduce the total 
systematic uncertainty of the polarimeter measurements to $\deltaP=0.2\percent$.
In the future, the calibration system should be tested with the full prototype
gas Cherenkov detector.

% \end{fmffile}

% \begin{acknowledgement}
 \section*{Acknowledgement}
The implementation and tests of the LED driver would not have been possible without the great technical 
support by Mathias Reinecke and Bernd Beyer. We thank our DESY colleagues in the CALICE collaboration
for fruitful discussions about LED calibration systems.
The results presented here could not be achieved without the National Analysis Facility and 
we thank the NAF team for their continuous support. We thankfully acknowledge the support by 
the BMBF-Verbundforschung in the context of the project ``Spin-Management polarisierter 
Leptonstrahlen an Beschleunigern''and the DFG by Li/1560-1. 
% \end{acknowledgement}

\appendix
\begin{footnotesize}
\bibliographystyle{apsrev}
\newcommand{\reftitle}[1]{``#1,''}

\end{footnotesize}

\end{document}